\documentclass[sigconf]{acmart}  
\usepackage{hyperref}
\usepackage{pdfpages} % http://mirror.unl.edu/ctan/macros/latex/contrib/pdfpages/pdfpages.pdf
\usepackage{booktabs} 

\usepackage{lineno}
\usepackage[utf8]{inputenc}
\usepackage{amsmath}
\usepackage{graphicx}
\usepackage[colorinlistoftodos]{todonotes} % handig voor commentaar: gebruik \todo{}, zie ftp://ftp.fu-berlin.de/tex/CTAN/macros/latex/contrib/todonotes/todonotes.pdf
\usepackage{listings}
\usepackage{pdfpages}
\usepackage{tcolorbox}
\usepackage{float}
\usepackage{caption}
\usepackage{subcaption}
\renewcommand\footnotetextcopyrightpermission[1]{}

\begin{document}

\begin{titlepage}

\begin{center}
 
\textsc{\Large   Prediction of Cardiovascular Risk Factors from Retinal Fundus Images using CNNs }

\bigskip

\textsc{\large
submitted in partial fulfillment for the degree of master of science\\
\bigskip
Andrea Prenner\\
13325973\\
\bigskip
master information studies\\
data science \\
faculty of science\\
university of amsterdam\\
\bigskip
2021-08-04
}

\end{center}

% logos
\bigskip
\bigskip
\bigskip
%\begin{center}
%\mbox{\includegraphics[width=.2\paperwidth]{logo-uva.png} 
%\includegraphics[width=.2\paperwidth]{TitlePages/logos/ads.png}
%\includegraphics[width=.2\paperwidth]{TitlePages/logos/ads.png} % replace by the logo of your internship company or remove
%}
%\end{center}

\vfill

\begin{center}
\begin{tabular}{|l||ll|}
\hline
 & \textbf{Internal  Supervisor} & \textbf{External   Supervisor}  \\   
 \hline
\textbf{Title, Name} & Dr Clarisa Sánchez Gutiérrez&  Sipko van Dam, PhD\\
\textbf{Affiliation} &UvA & Ancora Health\\ 
\textbf{Email} & c.i.sanchezgutierrez@uva.nl& sipko@ancora.health \\
\hline
\end{tabular}
\end{center}

\bigskip

\begin{center}
\mbox{\includegraphics[width=.22\paperwidth]{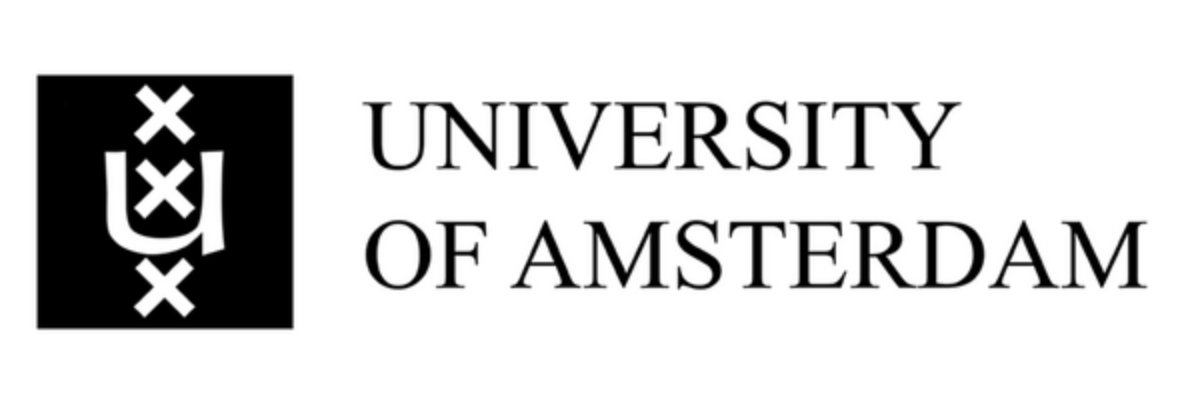} 
\includegraphics[width=.18\paperwidth]{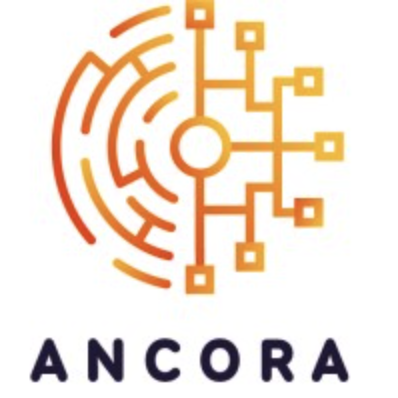}

}
\end{center}

\end{titlepage}

\title{Prediction of Cardiovascular Risk Factors \\
From Retinal Images using CNNs} % your title
\author{Andrea Prenner}

\begin{abstract}
Early detection of cardiovascular disease risk factors is essential to alter the course of the disease. Previous studies showed that deep learning can successfully be used to detect such risk factors from retinal images. This study uses convolutional neural networks (CNNs) to predict the cardiovascular disease risk factors age, BMI, smoking status, HbA1c, systolic blood pressure, diastolic blood pressure, gender and total cholesterol from retinal images from the UK Biobank data set. By applying contrast enhancement on the retinal images in the form of Gaussian filtering and deriving predictions on individual basis through the combination of left and right retinal image predictions, an increased prediction performance could be derived for the variables age (R2 score of 0.81) and systolic blood pressure (R2 score of 0.39) compared to previous studies using retinal images from the UK Biobank data set. Further, this is the first study that tries to predict HbA1c and total cholesterol from UK Biobank retinal fundus images. For these variables the models achieved an R2 score of 0.0579 for predicting HbA1c and an R2 score of 0.0157 for predicting total cholesterol. These results show that the value of deriving predictions for these two risk factors from retinal fundus images from the UK Biobank data set is limited.

\end{abstract}
\maketitle
\pagestyle{plain}

\section{Introduction}
%A clearly defined research problem and corresponding subquestions

According to the World Health Organization (WHO) cardiovascular diseases (CVD) represent the number one cause of deaths globally \cite{who}. 90\% of the risk of myocardial infarction stems from modifiable CVD risk factors, which indicates that CVD are preventable to a great extent. Therefore early detection of potential risk factors is essential to initiate effective counter measures such as lifestyle changes or pharmacotherapy \cite{cheong2017interventions}. Since most cardiovascular events occur with modest and consequently often unnoticed elevations of risk factors, risk detection is challenging \cite{dahlof2010cardiovascular}. Therefore, tools for early detection of individuals on the trajectory to develop CVD are essential in order to alter the course of the disease at an early stage by providing appropriate treatment \cite{dahlof2010cardiovascular}.

Traditional CVD risk screening requires a variety of variables derived from blood samples, the evaluation of the patient's history and physical measurements such as the BMI value. These variables are subsequently used to determine the patient's risk of experiencing a cardiovascular event within a specified time period and to ensure the appropriate treatment for the patients' at risk for such diseases \cite{poplin2018prediction}.

Since markers of CVD can often be observed in the eye, retinal images represent a way to predict these markers in a cheaper, quicker and non-invasive manner compared to traditional risk screening. Features in retinal images, like blood vessel calibre, tortuosity and bifurcation of the blood vessels, can be used to draw a conclusion about the health status of the cardiovascular system \cite{poplin2018prediction}. Specifically, Ding et al. \cite{ding2014retinal} showed that wider retinal venules and narrower arterioles indicate an increased risk of hypertension. McGeechan et al. \cite{mcgeechan2009meta} showed that these characteristics relate to an increased risk of CVD in general.

Recent research made use of deep learning in the form of convolutional neural networks (CNNs) to make predictions about CVD risk factors and further biomarkers \cite{poplin2018prediction}\cite{gerrits2020age}\cite{vaghefi2019detection}\cite{kim2020effects}\cite{zhang2020prediction} \cite{rim2020prediction}. By applying deep learning, which is a subgroup of machine learning, it is possible to learn appropriate predictive features from various sample images automatically rather than being obliged to hand-engineer features beforehand \cite{lecun2015deep}. This is an advantage compared to earlier work in automated classification of eye diseases, which relied on machine learning algorithms based on computing features specified by experts \cite{mookiah2013computer}. Deep convolutional networks have been used to diagnose diseases from medical images such as diabetic retinopathy and yielded an accuracy similar to a human expert \cite{gulshan2016development}. 

My research builds on the work of Poplin et al. \cite{poplin2018prediction}. Similar to the work of Poplin et al. \cite{poplin2018prediction} predictions for cardiovascular risk factors such as gender, smoking status, age, BMI, systolic blood pressure (SBP) and diastolic blood pressure (DBP) will be derived from retinal fundus images from the UK Biobank data set. Additional to those already analyzed variables, predictions will further be derived for total cholesterol and haemoglobin A1c (HbA1c) measurements (HbA1c was only predicted based on the EyePACS dataset in the study of Poplin et al.).

Since preprocessing retinal images with Gaussian filtering was found to yield a CVD risk factors prediction performance increase \cite{vaghefi2019detection} this contrast enhancement technique will be applied on the retinal images. To filter out images of poor-quality from the UK Biobank data set a model is trained on images whose image quality status was manually annotated. The CNN model structure that was used for the image quality assessment has been previously applied on retinal images from the Diabetic Retinopathy Image Database (DRIMDB) as well as the Brazilian Longitudinal Study of Adult Health (ELSA-Brasil) data set \cite{zago2018retinal}.

Since previous studies \cite{tajbakhsh2016convolutional} found that a deeper level of fine-tuning is essential for CNN models applied on medical images, I will test the difference in performance between a CNN model with pre-intialized ImageNet weights where only the weights of the last layers were updated in the training process and a fine-tuned CNN model with pre-intitialized ImageNet weights. 

Since the majority of individuals in the training set are from the british/irish population I will test whether the prediction performance is better for retinal images from individuals from the british/irish population than from individuals from other populations. If there cannot be a prediction difference observed between british/irish individuals and not british/irish individuals, it can be seen as indication that the model generalizes well also to other ethnicities. 

Further I will analyze if the combination of left and right retinal images per person leads to an increased prediction performance. Gerrits et al. \cite{gerrits2020age} observed a performance increase when information from four retinal fundus images per person were taken into account as opposed to solely one. 

Also I will investigate if gender and age have an influence on the prediction performance. Gerrits et al. \cite{gerrits2020age} found considerable differences in prediction performance between males and females for relative fat mass and testosterone. When testing the difference in performance between different age groups, the variables HbA1c and SBP showed considerable differences in performance.

Similar to the study of Poplin et al. \cite{poplin2018prediction} the inception-v3 model will be used in this research, which is widely adopted for medical image classification tasks \cite{christopher2018performance}.

The following research questions are addressed in this paper: 

\textbf{RQ1:} Does the prediction of the CVD risk factors total cholesterol and HbA1c by using the inception-v3 model with pre-trained weights, fine-tuned on retinal images from the UK Biobank data set, lead to a better prediction performance than a base line model? Since HbA1c and total cholesterol represent continuous variables the baseline model will be the average value.

\textbf{RQ2:} Does preprocessing the retinal fundus images in the form of Gaussian filtering and the application of an image quality assessment model to filter out images of inappropriate quality from the UK Biobank data set lead to an increase in performance compared to the study of Poplin et al.? 

\textbf{RQ3:} Does fine-tuning the model lead to a superior performance as opposed to training only the weights of the last added layers? 

\textbf{RQ4:} Can there be an increased prediction performance observed for retinal images from individuals belonging to the british/irish population? 

\textbf{RQ5:} Does the combination of CVD risk factor predictions from the left and right retinal fundus image per patient lead to an increased prediction performance? 

\textbf{RQ6:} Do the age level and sex have an influence on the CVD risk factor prediction performance?

\section{Related Literature}
%Overview of the state of the art of the literature 2. Clearly indicate how your approach is grounded in the literature
In the area of applying deep learning on retinal images, the work of Gulshan et al. \cite{gulshan2016development} was transformative since their inception-v3 model was able to detect diabetic retinopathy with high specificity (97.5\%) and sensitivity (93.4\%). 
More recently several studies with the focus on the prediction of cardiovascular risk factors from retinal fundus images via deep learning have been published \cite{poplin2018prediction}\cite{gerrits2020age}\cite{vaghefi2019detection}\cite{rim2020prediction}\cite{kim2020effects}\cite{zhang2020prediction}\cite{dai2020exploring}.

Poplin et al. \cite{poplin2018prediction} were one of the pioneers in the field by applying deep learning to predict CVD risk factors from retinal images. Specifically, they used the UK Biobank and EyePACS data set and applied an inception-v3 architecture with pre-trained weights where the weights were fine-tuned to retinal images in the training process. The predicted risk factors included age, gender, smoker status, HbA1c, BMI and blood pressure. For HbA1c prediction they used the EyePACS data set since at the time of their study HbA1c, glucose and cholesterol labels for retinal images were not available in the UK Biobank data set. As baseline comparison for continuous predictions they used the average value. The algorithm predicted age, blood pressure, BMI and HbA1c better than the baseline. However the predictions for BMI ($R^{2}$ of 0.13) and HbA1c ($R^{2}$ of 0.09) yielded a low $R^{2}$ indicating that the algorithm is not predicting these values with high precision. Conversely, the model achieved an area under the receiver operating curve (AUC) for the prediction of gender of 0.97 and for the prediction of smoking status of 0.71. Further they used the model to make predictions for the onset of major adverse cardiovascular events (MACE) within the next five years. Here the model received an AUC of 0.7. When applying the model of Poplin et al. \cite{poplin2018prediction} on a data set based on Asian patients, the prediction performance for CVD risk factors was similar across all data sets indicating that the model is generalizable to Asian populations (even though EyePACS and UK Biobank data set consists mainly of white individuals) \cite{ting2018eyeing}.

Rim et al. \cite{rim2020prediction} further expanded the study of Poplin et al. \cite{poplin2018prediction} by predicting additional biomarkers like body composition measurements and creatinine. They used a VGG16 model architecture where the weights were trained from scratch. The model predicted sex (AUC of 0.96), age ($R^{2}$ of 0.83), blood pressure (SBP $R^{2}$ of 0.31, DBP $R^{2}$ of 0.35), composition indices (Body muscle mass $R^{2}$ of 0.52, height $R^{2}$ of 0.42, bodyweight $R^{2}$ of 0.36, percentage body fat $R^{2}$ of 0.23, BMI $R^{2}$ of 0.17), serum creatinine concentration ($R^{2}$ of 0.38) and haematological parameters (Haematocrit $R^{2}$ of 0.57, Haemoglobin $R^{2}$ of 0.56, Red blood cell count $R^{2}$ of 0.45) well from retinal images. Even though the prediction performance of those variables was good on the internal test set, the models did not generalize well since when applied on external test sets the model yielded a considerably worse performance. Especially when tested on the UK Biobank data set the performance decreased substantially (sex AUC of 0.8, age $R^{2}$ of 0.51, height $R^{2}$ of 0.08, bodyweight $R^{2}$ of 0.04, BMI $R^{2}$ of 0.01, DBP $R^{2}$ of 0.16, SBP $R^{2}$ of 0.16, Haematocrit $R^{2}$ of 0.09, Haemoglobin $R^{2}$ of 0.06).  

Vaghefi et al. \cite{vaghefi2019detection} focused on the prediction of smoking status from retinal images. By using a self designed CNN, training the weights from scratch and applying Gaussian filtering on the retinal images, they could achieve an AUC of 0.86 with specificity of 93.87\% and sensitivity of 62.62\%. 

Zhang et al. \cite{zhang2020prediction} used an inception-v3 model with pre-trained weights to predict hypertension, hyperglycemia and dyslipidemia from retinal images. The model achieved an accuracy of 78.7\% in detecting hyperglycemia with an AUC of 0.880, an accuracy of 68.8\% in detecting hypertension with an AUC of 0.766 and an accuracy of 66.7\% in detecting dyslipidemia with an AUC of 0.703. The generalization of the model's prediction performance further needs to be evaluated on external data sets since they used a relatively small data set of 1,222 retinal images.

Dai et al. \cite{dai2020exploring} used retinal fundus images to predict hypertension. They designed their own CNN model and trained the weights from scratch. By applying an automated segmentation method which extracts retinal vessels of every fundus image and using these segmented images as inputs, the model yielded an improved performance with an accuracy of 60.94\% and an AUC of 0.6506. Nevertheless, the specificity was lower when using the segmented images given that information about non-hypertension is generated by other parts of the image not only blood vessels. Since the segmented images only contain blood vessels they are more likely to be classified as high blood pressure which leads to a higher recall but lower specificity. 

Also, various studies focused on investigating factors that might have an influence on the prediction performance of CVD risk factors. Gerrits et al. \cite{gerrits2020age} investigated the influence of sex and age on the prediction of various cardiovascular risk factors. For their research they used retinal images from the Qatar Biobank and used the MobileNet-V2 architecture with pre-trained weights as basis for their model. Gender had an impact on the prediction performance of testosterone and relative fat mass, whereas for the other variables age, DBP, SBP and HbA1c no considerable performance difference could be derived. Concerning the prediction impact of age Gerrits et al. \cite{gerrits2020age} could find that age has a significant impact on HbA1c and SBP prediction and some influence on testosterone and relative fat mass.

Poplin et al. \cite{poplin2018prediction} tested the influence of diabetic retinopathy (DR) severity on the model performance where they found no significant differences in the prediction performance of age, gender and HbA1c between the different DR severity levels. 

%Rim et al. \cite{rim2020prediction}, who trained their model on Asian population data, tested the influence of ethnic groups on the prediction performance. In their study they applied their model on each ethical group of the Singapore Epidemiology of Eye Diseases (SEED) study (Chinese, Indian, Malay) and the UK Biobank data (White, Non-White). They showed that prediction performance for systemic biomarkers differs by ethnicity. An explanation for the difference could be that the profiles of biomarkers differ across ethnicities. This finding contrasts the finding of Ting et al. \cite{ting2018eyeing} who found that the prediction performance of the model of Poplin et al. \cite{poplin2018prediction} was similar when applied on Asian population data sets. 

Kim et al. \cite{kim2020effects} predicted age and sex from retinal fundus images based on the retinal fundus images from the Seoul National University Bundang Hospital Retina Image Archive (SBRIA). They used a ResNet-152 architecture with pre-trained weights. The model yielded an $R^{2}$ of 0.92 for predicting age, which represents an increase compared to the study of Poplin et al. \cite{poplin2018prediction} ($R^{2}$ of 0.74 on the UK Biobank data set, $R^{2}$ of 0.82 on the EyePACS data set). They showed that the model for age prediction achieved the best results for participants in the age range of 20 to 40 years and that the accuracy in age prediction declined for subgroups with participants over 60 years. This could imply that the ageing process that is observable in retinal fundus images might saturate at about 60 years. The model yielded an AUC of 0.96 for predicting sex. The sex prediction performance showed no deterioration with increasing age. The generalization ability of their prediction model still needs to be evaluated since they did not test the model on external data sets. 

\section{Methodology}

My research will apply deep learning in the form of a CNN with pre-trained ImageNet weights that are fine-tuned in the training process to predict the CVD risk factors SBP, DBP, age, gender, BMI, smoker status, HbA1c and total cholesterol.

\subsection{Data}
The retinal fundus images from the UK Biobank data set will be used to train and validate the model. The UK Biobank is a data archive providing human health information across disease, health and demographics \cite{zhou2019predictive}. The available data set comprises 67,122 left eye and 67,230 right eye retinal fundus images. For 67,120 individuals left and right retinal fundus images are available. %The ethnicity groups range from xxx. 

For every individual various blood biomarkers and physical measurements are available. For some biomarkers, e.g. systolic blood pressure, multiple measurements from one time of measurement are available. In these cases the average of the measurements is taken.

\subsection{Transfer Learning Model}
Two models are generated in the course of this research: One model to assess the quality of the UK Biobank retinal images and classify them into good and bad quality images and one model to predict CVD risk factors from the good quality retinal images.  
For both models, the image quality assessment model and the CVD risk factor prediction model, the inception-v3 model architecture will be used. The inception-v3 model has been introduced by Szegedy et al. \cite{szegedy2016rethinking}. The techniques used in this model are called inception modules, which represent different sizes of convolutions and pooling applied separately on the same input and the resulting feature maps being concatenated and forwarded to the next inception module. The inception-v3 model is 42 layers deep \cite{sam2019offline} and has the advantage of performing well even under strict constraints of computational budget and on memory \cite{szegedy2016rethinking}. The inception-v3 model yields great performance when fine tuned on medical images \cite{esteva2017dermatologist} \cite{gulshan2016development}, outperforms other models when used on eye OCT images (only outperformed by the DenseNet-201 model, which needed a substantially higher training time as opposed to the inception-v3 model)\cite{islam2019identifying} and is widely adopted for medical image classification tasks \cite{christopher2018performance}. Further it has been successfully applied for classification of retinal fundus images \cite{poplin2018prediction} \cite{gulshan2016development} \cite{li2018automated}. 

In both models transfer learning will be applied, where the CNN model with pre-trained weights from a natural image data set is fine tuned to retinal images. The pre-trained weights are taken from an inception-v3 model that has been trained on the ImageNet data set. The ImageNet data set comprises 1.2 million 256x256 images categorized in 1,000 object class categories \cite{shin2016deep}. Since the pre-trained network had learned various features from the ImageNet data, transfer learning shows faster performance than learning the weights from scratch \cite{yu2017deep}.
To adjust the pre-trained model to the medical domain, fine-tuning is applied by using the pre-trained weights as initialization and updating them in the training process to capture domain-specific features. Tajbakhsh et al. \cite{tajbakhsh2016convolutional} showed that deeper fine-tuning yielded comparable or even superior performance to CNNs trained from scratch, whereas shallow fine tuning led to a performance inferior to CNNs trained from scratch. This performance gap between deeply fine tuned CNNs and CNNs trained from scratch increased as the size of training sets was reduced. Further the depth of fine-tuning is essential to yield accurate image classifiers. They found that a deeper level of tuning is essential for medical imaging applications. 

As optimization technique the Adam optimizer \cite{kingma2014adam} is used in the training process of both models. The Adam optimizer is one of the most popular optimization algorithms for gradient descent \cite{bock2018improvement}. Using this optimizer instead of SGD has the advantage of faster convergence towards the minimum \cite{ruder2016overview} and is therefore computationally efficient, has little memory requirements and is well suited for problems that are large regarding data and/or parameters \cite{kingma2014adam}. The default settings of a learning rate of 0.001, a beta 1 of 0.9 and a beta 2 of 0.999 will be used. 

Further early stopping is applied where the training process is stopped if the validation loss has not been improving over the last 50 epochs \cite{gerrits2020age}. The model with lowest validation loss over all training epochs is saved. 

In the training process of both models, data augmentation is applied to the training set in order to reduce overfitting and enhance generalization of the model since data augmentation exposes the model to more aspects of the data \cite{perez2017effectiveness}. Data augmentation is a way to increase the training data by applying transformation techniques on the initial images \cite{zago2018retinal}. The random transformations applied on the training images involve rotation and vertically and horizontally flipping of the images.

The models are both developed and deployed with TensorFlow open-source software libraries.

\subsection{Image Quality Assessment Using Deep Learning}
Since poor-quality retinal images could lead to inaccurate medical diagnosis, an image filtering process is applied to filter out bad quality images from the UK Biobank data set. Previous image filtering methods for the UK Biobank data set by Welikala et al. \cite{welikala2016automated} involved blood vessel features to train an SVM classifier to assess image quality. Applying their model resulted in a classification of 26\% of the images in the UK Biobank as inadequate quality. MacGillivray et al. \cite{macgillivray2015suitability} found that only 36\% of the individuals in the UK Biobank data set had both retinal images in adequate quality for further analysis. In the work of Poplin et al. \cite{poplin2018prediction} 12\% of individuals in the UK Biobank were filtered out due to poor quality. Images where the circular mask in the retinal fundus image could not be detected or the images were of poor quality were filtered out. Zago et al. \cite{zago2018retinal} made use of deep learning to train a CNN model to classify retinal images into good and bad quality. For their analysis they used retinal images from the Diabetic Retinopathy Image Database (DRIMDB) as well as the Brazilian Longitudinal Study of Adult Health (ELSA-Brasil) data set. They used the inception-v3 model as basis and had three training cycles. In a first model they trained the model's weights from scratch. In a second model they used pre-trained weights from the ImageNet data set and trained only the last added fully connected layers. Last, they used pre-trained weights and fine-tuned them over all layers. They found that the fine-tuned model with pre-trained weights yielded the best performance. The quality assessment model I applied is based on the model of Zago et al. \cite{zago2018retinal}. 

For the image quality assessment in this paper, a binary classification CNN model is trained in order to distinguish good from bad quality images. The data set used to train the model encompasses 1,289 manually annotated images. In order to prevent class imbalance in the training data set, a similar amount of bad (582 images) and good (707 images) quality images has been labelled. Further an equal amount of left (642) and right (647) eye retinal images has been labelled. The images are cropped around the circular mask in order to remove black background pixels. Then the images are resized to a size of 587x587. Similar to the study of Poplin et al. \cite{poplin2018prediction} this image size will later be used for the CVD risk factors' prediction as well. 
Preprocessing in the form of normalization of all pixel values from the interval [0,255] to [-1,1] is applied.  
Similar to the study of Zago et al. \cite{zago2018retinal} the inception-v3 architecture will be used with pre-initialized weights from training on the ImageNet dataset \cite{russakovsky2015imagenet}. Since the model was trained to classify 1,000 classes, the last two layers are removed and a layer with 1,024 neurons, a layer with 512 neurons and an output layer with one neuron, are added to the model. The output layer will use a sigmoid activation function to derive a probability value, whereas the outputs of the other two fully connected layers are activated by a Rectified Linear Unit (ReLU) activation function. 

To prevent overfitting, dropout will be applied after the first two added layers with setting the keep rate of each unit to 0.5, which seems to be close to optimal for a wide range of tasks and networks. The term dropout refers to dropping out units in the neural network. Specifically, units and their incoming and outgoing connections are temporarily removed from the network. The choice of which units to drop is random. Each unit is retained with a fixed probability p, which is independent of other units \cite{srivastava2014dropout}.

The threshold to distinguish between good and bad quality is arbitrarily set to 0.5, where images with a probability value above 0.5 are classified as good quality image and below 0.5 as bad quality image. The loss function that is used in the optimization process is the Binary Cross Entropy Loss function. 

The 1,289 annotated images were split into training (60\%), validation (20\%) and test set (20\%), where it was ensured that the same proportion of good and bad quality images was in each subset. The model was trained in two ways. First, only the weights of the added layers were trained and the pre-trained ImageNet weights were used in the other layers. Second, all layers' weights were trained and the pre-trained weights were used as initialization. 

Before the image quality assessment model was applied on the UK Biobank retinal images, images from individuals where not both, left and right, retinal images were available, were discarded. 
On the remaining images from 67,120 individuals, the quality classification model was applied. Then again images were filtered out where only one good quality image per individual could be derived. 
The filtered data set which was used for further analysis only contains images, where left and right retinal images of appropriate quality per individual were available. 

%First check if right and left eye image are available and then apply model. -> Has to be applied to fewer images. 

\subsection{Preprocessing}
The images with adequate quality were preprocessed by scale-normalizing them through cropping the image so that the circular mask is fully in the image and resizing them to 587x587 pixels. The circular mask was detected by applying a thresholding technique. The images were preprocessed following the retinal image preprocessing techniques of Graham \cite{graham2015kaggle} to enhance the contrast in the images and which have been successfully applied in previous studies \cite{gonzalez2019iterative}. 

\begin{center}
    \includegraphics[scale=0.195]{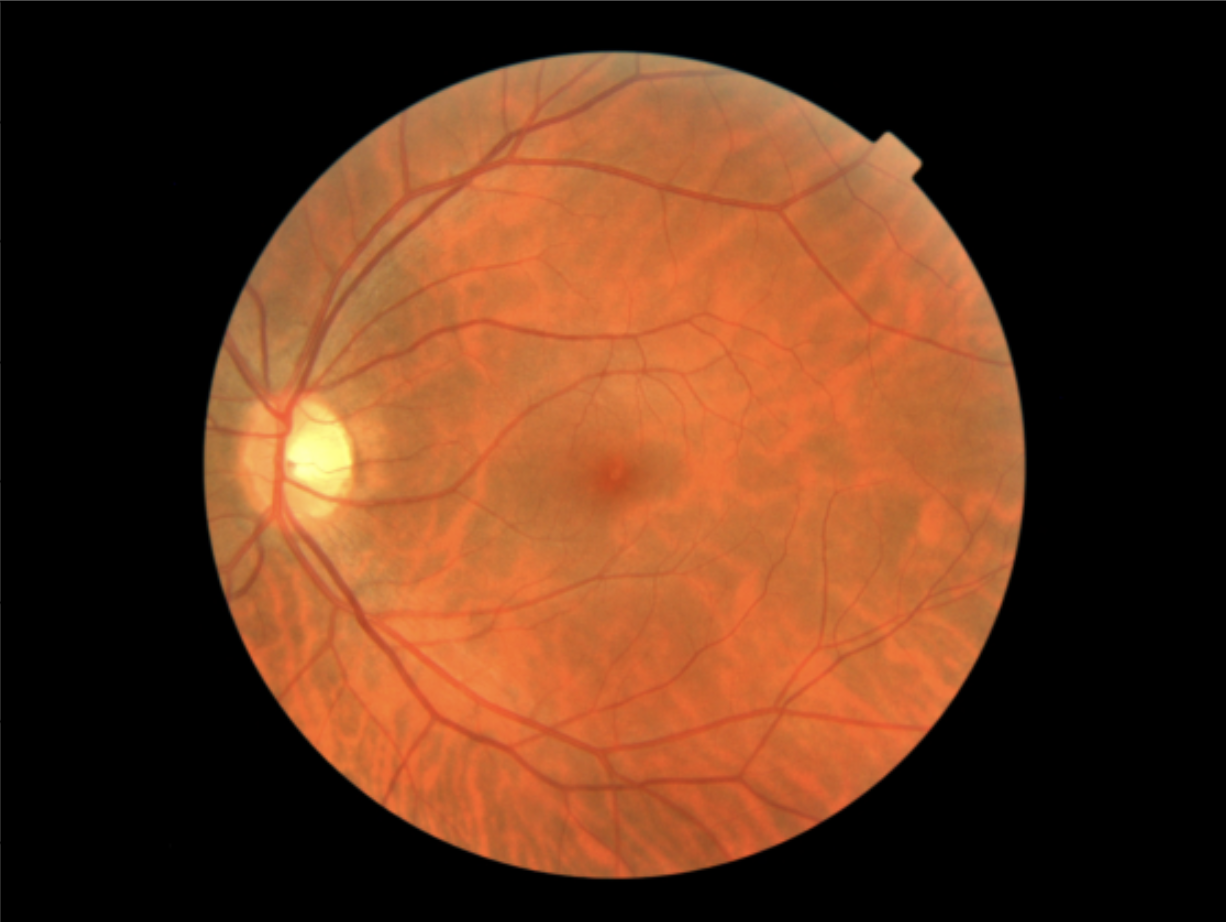}
    \includegraphics[scale=0.15]{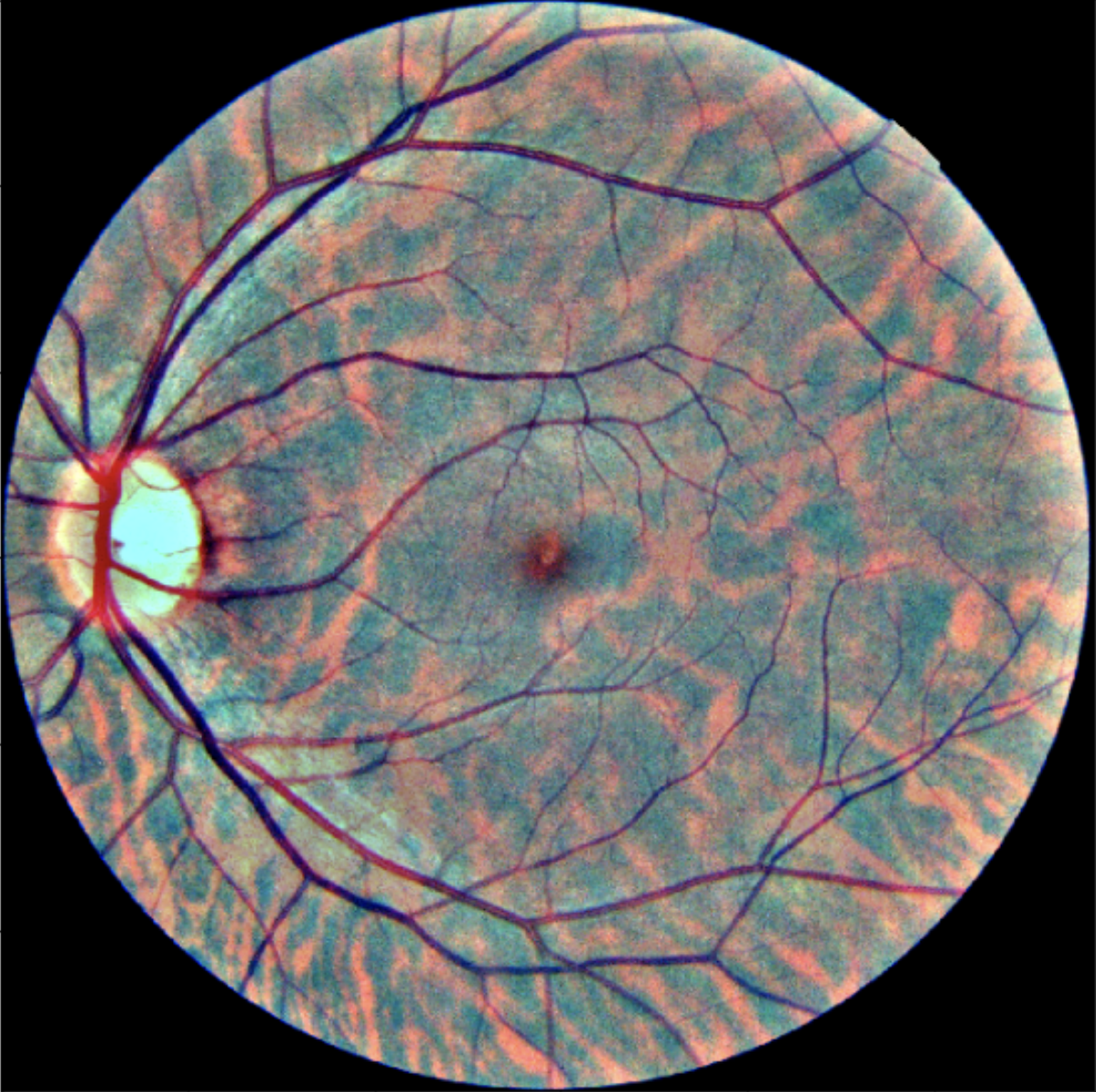}\\
    Figure 1. Original image (left) and preprocessed image (right).
\end{center}

The image pixel values were then normalized to a range of [-1,1]. 

The images were split in train, validation and test set with a split of 60\%, 20\% and 20\%. To have unbiased results when testing the performance difference between predictions on image basis as opposed to predictions on individual basis (combination of left and right retinal image predictions), it is ensured that the two retinal images per person are within one subset.

\subsection{CVD Risk Factor Prediction Model}
The inception-v3 model architecture was used for the prediction of the CVD risk factors. Similar to the study of Gerrits et al. \cite{gerrits2020age}, who used the MobileNet-V2 model architecture, a global average pooling layer and two layers were added to the inception-v3 model. The first fully connected layer has 512 neurons that are each activated with a ReLU activation function. For the regression tasks the output layer has one neuron with a linear activation, for the classification tasks the output layer also has one unit, but a sigmoid activation. Again after the first fully connected layer dropout was applied with a p value of 0.5. 

The prediction model for the continuous variables (BMI, age, SBP, DBP, total cholesterol, HbA1c) was trained by optimizing the mean absolute error loss function. 

Similar to the study of Poplin et al. \cite{poplin2018prediction} the smoking status was binarized into current smoker yes or no. Therefore the predictions of gender and smoking status represent binary classification problems where the binary cross entropy loss is optimized in the course of the training process. 

Similar to the study of Poplin et al. \cite{poplin2018prediction} and Gerrits et al. \cite{gerrits2020age} a batch size of 32 was used. 

Training a CNN on an imbalanced data set in which one class contains the majority of the data instances and the other class contains far fewer instances leads to classifiers that are often biased towards the majority class, which results in a higher misclassification rate of the minority class \cite{lopez2013insight}. In the case of the categorical variable smoking status the data is highly imbalanced towards the non-smoking class. In this case a weighted loss function was introduced that gives higher weight to the loss of the minority class (smoking class). Specifically, the class weights were set inversely proportional to the class frequency. Apart from proportionally weighting the loss, undersampling the majority class or oversampling the minority class are further options to deal with class imbalance. Therefore, besides the weighted loss approach, the approach of undersampling the majority class was evaluated in this research. For this approach the proportion of smoking and non-smoking individuals was balanced out by randomly sampling 1,500 non-smoking individuals from the training set. This resulted in left and right eye scans of 1,500 non-smoking and 1,470 smoking individuals that were used to train the model. 

For each of the variables a separate prediction model was trained where the loss is optimized for each model individually. If an image had a missing value for the variable that was aimed to be predicted, it was discarded in the training process.

Further the impact on performance by using information from the left and right retinal image per person was evaluated. Specifically, in the case of regression tasks the continuous prediction output values for the left and right retinal images were averaged whereas in the case of classification tasks the output probabilities were averaged. Gerrits et al. \cite{gerrits2020age} observed a performance increase when averaging the output of the model for four retinal images per person. 

%\begin{center}
%\mbox{\includegraphics[width=.4\paperwidth]{two_inputs.png}}
    % \medskip
    % \tiny
    % \centering
     %\medskip
        %  Figure 3: Averaging Output of Two Input Images
%\end{center}

\section{Results}

To evaluate the model performance for continuous variable predictions, the coefficient of determination ($R{2}$ score) and the mean absolute error (MAE) are calculated. As baseline comparison the mean value of each variable is used.  

For binary classifications, the area under the ROC curve (AUC) is calculated. As baseline comparison an AUC score of 0.5 is used. The AUC equals 1 if the classifier can perfectly distinguish between negative and positive examples whereas a classifier with random guessing yields an AUC of 0.5 \cite{whitehill2016exploiting}. Since in the case of smoking status prediction the data set is highly imbalanced, the performance was further evaluated by analyzing the area under the precision-recall curve (AUC-PR).  

To evaluate the statistical significance of the prediction results, non-parametric bootstrapping is applied on the test set to estimate confidence intervals. 
In non-parametric bootstrapping resampling from the data set with replacement is applied. For each sample the statistic of interest is estimated. N replicated samples are generated that are used to estimate the distribution of the statistic of interest \cite{nixon2010non}. Similar to Gerrits et al. \cite{gerrits2020age} the statistics are calculated from random samples that are obtained from the test set with replacement containing as much samples as the test set. From the resulting distribution 95\% confidence intervals are calculated. 

\subsection{Image Quality Model}
In the non-parametric bootstrapping procedure the number of images randomly sampled with replacement equals 168, the size of the test set. The sampling process is repeated 1,000 times. As a rule of thumb 1,000 bootstrap replicates are needed to estimate confidence intervals \cite{nixon2010non}. For each of the samples the AUC, precision, recall and accuracy score is derived. Thereby a distribution of the AUC, precision, recall and accuracy score could be derived from which the 95\% confidence interval (2.5 and 97.5 percentiles) is reported. The recall, precision and accuracy score for the classification model are each derived by setting the threshold to 0.5.

\begin{table}[H]
\centering
 \scriptsize
   \label{tab:table1}
    \begin{tabular}{c|c|c|c|c}
      \textbf{Model}&\textbf{Accuracy}&\textbf{AUC}&\textbf{Precision}&\textbf{Recall}\\
    \hline
    Inception-v3 pre-trained weights & 0.9109  & 0.9707  &	0.8839  & 0.9648  \\
    training of last FC layers & (0.88, 0.94) & (0.95, 0.99) & (0.83, 0.93)  & (0.93, 0.99) \\
    \hline
      
    Inception-v3 pre-trained weights 	& 0.9341 & 0.9780 & 0.9139 & 0.9718 \\
    fine-tuning of all layers & (0.90, 0.96) & (0.96, 0.99) & (0.87, 0.96) & (0.94, 0.99)
    
    \end{tabular}
     \medskip
     \tiny
     \centering
     \medskip
    \\Table 1: Performance of image quality assessment model with pre-trained ImageNet weights and \\
    with fine-tuning over all layers (Confidence Interval)
\end{table}

\begin{center}
    \includegraphics[scale=0.4]{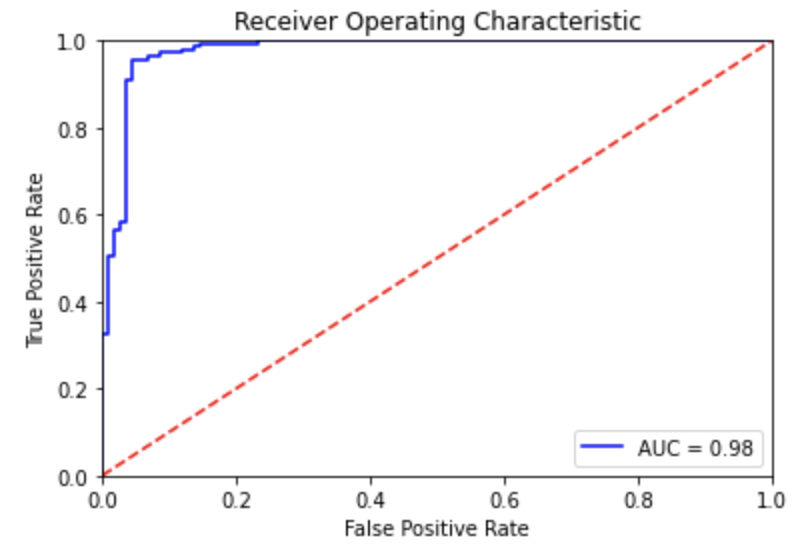}\\
    Figure 2. ROC curve of fine-tuned image quality assessment model
\end{center}

Since the model with fine-tuned weights over all layers outperformed the model where only the added layers' weights were trained, this model was applied to filter out images of appropriate quality from the UK Biobank data set for further analysis. 
The image quality assessment model was applied on the images where both left and right retinal image per individual were available. From the 67,120 left eye retinal images 35,991 were classified as good quality (53.62\%). From the 67,120 right eye retinal images 39,179 were classified as good quality (58.37\%). Based on the chi-square test of independence the percentage of bad quality images is not independent from left or right retinal image subgroup since the p-value is below 0.05 (p-value equals 9.51e-54). %Does that mean that statistically signficant for left images more images are bad quality? 

After discarding individuals where not both left and right eye retinal images were of appropriate quality, the final data set for further analysis resulted in 62,026 left and right retinal images from 31,013 individuals.

\subsection{CVD Risk Factor Prediction Model}

In table 2 an overview of the characteristics of the variables in the development set and the test set is given. The development set encompasses the training and validation set and is used to build the CVD risk factor prediction models. The test set is used to test the final models on unseen data. The minimum age in the data set used for the analysis is 39.67 years. Therefore when testing the influence of age level on the prediction performance this can only be tested for older age levels as opposed to the study of Gerrits et al. \cite{gerrits2020age}, who also tested the influence of age below 39 on the prediction performance. 

\begin{table}[H]
\centering
 \scriptsize
   \label{tab:table1}
    \begin{tabular}{l|l|l|l|l}
     \textbf{} & \textbf{Development Set} & \textbf{Test Set}\\
     \hline
    Number of individuals & 24,810 & 6,203   \\
    \hline
    Number images 	& 49,620 & 12,406 \\
    \hline
    Ethnicity (\%British/Irish)	& 91.10 & 91.17 \\
    \hline
    Gender (\%male) & 42.87, n = 23,914 & 42.87, n = 6,203 \\
    \hline
    Smoking Status (\%Current) 	& 7.88, n = 21,635 & 8.17, n = 5,401\\
    \hline
    Age (Mean, SD) 	& 56.04 (8.24), n = 24,810 & 56.11 (8.26), n = 6,203 \\
    \hline
    BMI (Mean, SD) 	& 27.15 (4.70), n = 24,704 & 27.06 (4.62), n = 6,172 \\
    \hline
    SBP (Mean, SD) 	& 136.23 (18.25), n = 24,789 & 136.27 (18.18), n = 6,196 \\
    \hline
    DBP (Mean, SD) 	& 81.72 (10.01), n = 24,789 & 81.76 (10.00), n = 6,197 \\
    \hline
    HbA1c (Mean, SD) 	& 35.61 (6.37), n = 22,550 & 35.47 (5.94), n = 5,612 \\
    \hline
    Total Cholesterol (Mean, SD) & 5.72 (1.12), n = 22,859 & 5.70 (1.13), n = 5,689 \\
    \hline
    %LDL Cholesterol (Mean, SD) 	& 3.56 (0.85), n = 22,818 & 3.54 (0.86), n = 5,673 \\
    %\hline
    %HDL Cholesterol (Mean, SD) 	& 1.49 (0.39), n = 21,800 & 1.49 (0.39), n = 5,673 \\
    %\hline
    
    \end{tabular}
     \medskip
     \tiny
     \centering
     \medskip
    \\Table 2: Data characteristics of final UK Biobank that is used for the analysis, n equals the number of individuals for whom the measurement was available
\end{table}

To evaluate if an increased prediction performance is given by training all layers' weights as opposed to only training of the weights of the last layers, for the variables gender and age two models with these two different training approaches were generated. Even though the models with training of only the last layers outperformed the baseline models, they were outperformed by the fine-tuned models for both the classification and the regression task. For example for age the prediction model without fine-tuning achieved an $R^{2}$ score of 0.5520 with a mean absolute error (MAE) of 4.6029 years (see table 4) whereas with fine-tuning the $R^{2}$ score equaled 0.8070 and the MAE equaled 2.7711 years (see table 6). For the prediction of gender the AUC score increased from 0.6701 without fine-tuning (see table 3) to 0.9533 with fine-tuning (see table 5). Since this confirms previous findings that for medical images fine-tuning plays an essential role \cite{tajbakhsh2016convolutional}, for the other variables in the analysis (SBP, DBP, BMI, Cholesterol, HbA1c, Smoking status) only the fine-tuned models were generated. Even though the fine-tuned models achieved a superior performance, it needs to be highlighted that training only the last layers is significantly faster since fewer parameters need to be updated in the training process and therefore they need less computational power to be trained. 

Since for all variables the combination of predictions from the left and right retinal images resulted in an increased prediction performance, only the prediction performance after combination of left and right eye predictions is displayed in the tables below. The statistics concerning the prediction performance without combination of left and right eye information can be found in the appendix (section A, B). For example by combining the information from both retinal scans the prediction performance of age increases from an $R^{2}$ score of 0.7714 (prediction performance on image basis - see table 15 in appendix) to 0.8070 (prediction performance on individual basis - see table 6).

\begin{table}[H]
\centering
 \scriptsize
   \label{tab:table1}
    \begin{tabular}{c|c|c}
     \textbf{Risk Factor} &  \textbf{AUC} & \textbf{Baseline}\\
     \hline
    Gender & 0.6701 & 0.5 \\
    (95\% CI) & (0.66, 0.68) &   \\
    \hline
    
    \end{tabular}
     \medskip
     \tiny
     \centering
     \medskip
    \\Table 3: Gender prediction with inception-v3 model with training of weights of last layers
\end{table}

\begin{table}[H]
\centering
 \scriptsize
   \label{tab:table1}
    \begin{tabular}{c|c|c|c|c}
     \textbf{Risk Factor} &  \textbf{MAE} & \textbf{R2 Score} & \textbf{Baseline MAE} & \textbf{Baseline R2}\\
  \hline  
    Age & 4.6029 & 0.5220 & 7.1681 & 0  \\
    (95\% CI) & (4.54, 4.66) & (0.51, 0.53) & &\\
    \hline
    
    \end{tabular}
     \medskip
     \tiny
     \centering
     \medskip
    \\Table 4: Age prediction with inception-v3 with training of weights of last layers
\end{table}

The final performance results of the fine-tuned model on the test set are displayed in table 5 for categorical risk factors and in table 6 for continuous risk factors. Both the prediction models for the categorical variables and the continuous variables performed better than the baseline. For prediction of gender the model achieved an AUC of 0.9533, which is lower than the AUC score of 0.97 achieved by the model of Poplin et al. \cite{poplin2018prediction}, the AUC score of 0.96 from the model of Gerrits et al. \cite{gerrits2020age} and the AUC score of 0.96 predicted by the model of Kim et al. \cite{kim2020effects}. The AUC-PR score equals 0.9458. 

Concerning the prediction of smoking status, the performance of the model with a weighted loss function achieved a superior performance compared to the model that used undersampling to counteract class imbalance (AUC of 0.6856 vs. AUC of 0.5666), therefore only the results of the model with weighted loss are displayed in the tables below. The performance results of the model with undersampling can be found in the appendix (section B). Compared to previous studies the AUC of 0.6856 is slightly lower than the AUC of 0.71 yielded by the model of Poplin et al. \cite{poplin2018prediction} and substantially lower than the AUC of 0.78 and 0.86 achieved by the models of Gerrits et al. \cite{gerrits2020age} and Vaghefi et al.  \cite{vaghefi2019detection} respectively. The AUC-PR of smoking status prediction equals 0.2298. 

\begin{table}[H]
\centering
 \scriptsize
   \label{tab:table1}
    \begin{tabular}{c|c|c|c}
     \textbf{Risk Factor} &  \textbf{AUC} &  \textbf{AUC-PR} &  \textbf{Baseline} \\
    \hline
    Gender & 0.9533 & 0.9458 & 0.5 \\
    (95\% CI) &  (0.95, 0.96) & (0.94, 0.95) & \\
    \hline
    Smoking &  0.6856 &  0.2298 & 0.5 \\
    (95\% CI) &  (0.67, 0.70)& (0.21, 0.26) &  \\
    \hline

    \end{tabular}
     \medskip
     \tiny
     \centering
     \medskip
    \\Table 5: Categorical risk factor prediction with inception-v3 model with fine tuning of all layers
\end{table}

For predicting age the model achieved an $R^{2}$ score of 0.8070 with a MAE of 2.7711 years. This represents an increase compared to the study of Poplin et al. \cite{poplin2018prediction}, who achieved an $R^{2}$ score of 0.74 for the prediction of age. Still the coefficient of determination is lower than the model of Gerrits et al. \cite{gerrits2020age}, who achieved an $R^{2}$ score of 0.85 with a MAE of 3.26 years on the prediction of age. Also the model of Kim et al. \cite{kim2020effects} achieved a higher $R^{2}$ score of 0.92 for predicting age. 

Systolic blood pressure (SBP) could be predicted with a MAE of 11.0036 mmHg and an $R^{2}$ score of 0.3927. This performance is slightly higher than the model of Poplin et al. ($R^{2}$ score of 0.36), but slightly lower than the model of Gerrits et al. ($R^{2}$ score of 0.4). Even though SBP could be slightly better predicted than by the model of Poplin et al., the prediction performance of DBP is slightly lower with an $R^{2}$ score of 0.2910 as opposed to 0.32 from the model of Poplin et al. \cite{poplin2018prediction}. 
BMI could be predicted with a MAE of 3.2119 kg/$m^{2}$ and an $R^{2}$ score of 0.0960, which represents a lower performance than the model of Poplin et al. ($R^{2}$ score of 0.13) and Gerrits et al. ($R^{2}$ score of 0.13). The model for predicting HbA1c achieved an $R^{2}$ score of 0.0579. This is lower than the model of Poplin et al. which achieved an $R^{2}$ score of 0.09. Here it has to be highlighted that for the prediction of HbA1c in the model of Poplin et al. \cite{poplin2018prediction} only images from the EyePACS and not from the UK Biobank data set were used. The predictions for HbA1c by the model of Gerrits et al. was substantially higher with an $R^{2}$ score of 0.34. Even though the prediction performance of cholesterol is a little better than the baseline model, the performance is still low with an $R^{2}$ score of 0.0157. Also the prediction performance is a lower than the performance by Gerrits et al. \cite{gerrits2020age}, whose model achieved an $R^{2}$ score of 0.03. The prediction of total cholesterol was not included in the study of Poplin et al. \cite{poplin2018prediction}. 

\begin{table}[H]
\centering
 \scriptsize
   \label{tab:table1}
    \begin{tabular}{c|c|c|c|c}
    \textbf{Risk Factor} &  \textbf{MAE} & \textbf{R2 Score} & \textbf{Baseline} & \textbf{Baseline}\\
    \textbf{} &  \textbf{} & \textbf{} & \textbf{MAE} & \textbf{R2}\\
    \hline
    Age & 2.7711 & 0.8070 & 7.1681 & 0  \\
    (95\% CI) & (2.73, 2.81) & (0.80, 0.81) & &\\
    \hline
    SBP & 11.0036 & 0.3927 & 14.3644  & 0  \\
    (95\% CI) & (10.85, 11.16) & (0.38, 0.41) & &\\
    \hline
    DBP & 6.6689 & 0.2910 & 7.9256 & 0  \\
    (95\% CI) & (6.58, 6.76) & (0.28, 0.30) & &\\
    \hline
    BMI & 3.2119 & 0.0960 & 3.4936 & 0  \\
    (95\% CI) & (3.16, 3.27) & (0.08, 0.11) & &\\
    
    \hline
    Cholesterol & 0.8858 & 0.0157 & 0.8942 & 0  \\
    (95\% CI) & (0.87, 0.90) & (0.01, 0.02) & &\\
    \hline
    HbA1c & 3.3937 & 0.0579 & 3.6141 & 0  \\
    (95\% CI) & (3.31, 3.48) & (0.05, 0.07) & & \\
    \hline
    
    \end{tabular}
     \medskip
     \tiny
     \centering
     \medskip
    \\Table 6: Continuous risk factor prediction with inception-v3 model with fine tuning of all layers
\end{table}

Since the prediction performance for some variables like age and SBP outperformed the model of Poplin et al. \cite{poplin2018prediction} whereas for other CVD risk factors an inferior performance was observed, I further analyzed if the Gaussian filtering technique only yields an advantage for some variables (age, SBP) whereas for other variables with worse performance like gender other preprocessing techniques might need to be considered. Therefore for predicting gender another fine-tuned model was trained with unprocessed images (only cropping around circular mask has been applied as preprocessing). The prediction results can be found in table 7. Without preprocessing the input images the prediction performance decreased from an AUC of 0.9533 to an AUC of 0.9411. Therefore the Gaussian filtering technique is likely enhancing the prediction performance for all risk factors.

\begin{table}[H]
\centering
 \scriptsize
   \label{tab:table1}
    \begin{tabular}{c|c|c}
     \textbf{Risk Factor} &  \textbf{AUC} & \textbf{Baseline}\\
     \hline
    Gender & 0.9411 &  0.5 \\
    (95\% CI) & (0.94, 0.95) &   \\
    \hline
    
    \end{tabular}
     \medskip
     \tiny
     \centering
     \medskip
    \\Table 7: Gender prediction with fine-tuned inception-v3 model with unprocessed input images \\ Prediction performance on individual basis
\end{table}

Further to test if there is a superior prediction performance for retinal images from the british/irish population, which is the dominant one in the UK Biobank data set, the predictions for retinal images from the british/irish population are compared to predictions for retinal images not from this specific population within the test set (see table 8). For the variables gender, SBP and HbA1c no statistical significant difference between the prediction performances of these two population groups could be observed. For the variables smoking status, age, BMI and total cholesterol the model achieved a higher prediction performance for the british/irish population group, whereas for the variable DBP the model achieved a lower performance for the british/irish population group. Since only for some variables the prediction performance for british/irish individuals is better than for other ethnicities, it cannot be concluded that the models overall achieve better CVD risk factor predictions for british/irish individuals.

\begin{table}[H]
\centering
 \scriptsize
   \label{tab:table1}
    \begin{tabular}{c|c|c}
    \textbf{Risk Factor} &  \textbf{British/Irish} & \textbf{Not British/Irish}  \\
    \hline
    Gender  & 0.9538 & 0.9475  \\
    (95\% CI) & (0.9501, 0.9572) &  (0.9437, 0.9512)\\
    \hline
    Smoking  & 0.6900 & 0.6251 \\
    (95\% CI) & (0.6717, 0.7083) & (0.6087, 0.6400) \\
    \hline
    Age & 0.8063 & 0.7759 \\
    (95\% CI) & (0.7993, 0.8130) & (0.7666, 0.7844) \\
    \hline
    SBP & 0.3906 & 0.3910 \\
    (95\% CI) & (0.3769, 0.4033) & (0.3763, 0.4051) \\
    \hline
    DBP & 0.2883& 0.3195 \\
    (95\% CI) & (0.2750, 0.3017) & (0.3065, 0.3325) \\
    \hline
    BMI & 0.0991& 0.0670 \\
    (95\% CI) & (0.0882, 0.1101)  & (0.0545, 0.0788) \\
    \hline
    Cholesterol & 0.0159& 0.0032 \\
    (95\% CI) & (0.0106, 0.0212) & (-0.0025, 0.0086) \\
    \hline
    HbA1c & 0.0579& 0.0586 \\
    (95\% CI) & (0.0494, 0.0661) & (0.0489, 0.0690) \\
    \hline
    \end{tabular}
     \medskip
     \tiny
     \centering
     \medskip
     \\Table 8: Prediction performance of categorical risk factor prediction model over different ethnicities \\ For categorical variable prediction AUC shown, for continuous variables $R^{2}$ shown.
\end{table}

In table 9 the prediction performance for left and right retinal images are summarized. It can be derived that the prediction performance from left and right retinal images for all variables is not statistically significant different from each other since the confidence intervals are overlapping for each of the variables.
%Since the right retinal image group was found to have a statistically significant higher percentage of good quality images as opposed to the left retinal image group when applying the image quality assessment model (see section 4.1) , it could be that the remaining good quality retinal images from the right eyes are still in better quality than from the left eyes and therefore this results in this prediction performance difference. 

\begin{table}[H]
\centering
 \scriptsize
   \label{tab:table1}
    \begin{tabular}{c|c|c}
    \textbf{Risk Factor} &  \textbf{Left Retinal Image} & \textbf{Right Retinal Image}  \\
    \hline
    Gender  & 0.9307 & 0.9286 \\
    (95\% CI) & (0.9263, 0.9351) & (0.9239, 0.9331) \\
    \hline
    Smoking  & 0.6680  & 0.6754  \\
    (95\% CI) & (0.6500, 0.6860) & (0.6577, 0.6929) \\
    \hline
    Age & 0.7667 & 0.7761 \\
    (95\% CI) & (0.7583, 0.7754) & (0.7684, 0.7834) \\
    \hline
    SBP & 0.3203 & 0.3402 \\
    (95\% CI) & (0.3046, 0.3371) & (0.3248, 0.3564) \\
    \hline
    DBP & 0.2441 & 0.2557 \\
    (95\% CI) & (0.2292, 0.2597) & (0.2416, 0.2702) \\
    \hline
    BMI & 0.0775 & 0.0576 \\
    (95\% CI) & (0.0642, 0.0894) & (0.0439, 0.0709) \\
    \hline
    Cholesterol & 0.0069 & 0.0094 \\
    (95\% CI) & (0.0003, 0.0130) & (0.0032, 0.0154) \\
    \hline
    HbA1c & 0.0539 & 0.0569 \\
    (95\% CI) & (0.0452, 0.0628) & (0.0481, 0.0658) \\
    \hline
    \end{tabular}
     \medskip
     \tiny
     \centering
     \medskip
    \\Table 9: Prediction performance left and right retinal image \\ For categorical variable prediction AUC shown, for continuous variables $R^{2}$ shown.
\end{table}

To test the influence of gender on the prediction performance, the prediction performance between male and female subgroups was compared (see table 10). For the variables SBP, cholesterol and HbA1c a statistical difference in prediction performance between the gender groups could be reported. This is contrary to the findings of Gerrits et al. \cite{gerrits2020age}, who only found considerable performance differences for the variables testosterone and relative fat mass between the gender groups. 

\begin{table}[H]
\centering
 \scriptsize
   \label{tab:table1}
    \begin{tabular}{c|c|c}
     \textbf{Risk Factor} & \textbf{Male} & \textbf{Female}  \\
     \hline
    Age  & 0.8075 & 0.8064 \\
    (95\% CI)  & (0.8008, 0.8137) & (0.7993, 0.8132) \\
    \hline
    SBP  & 0.3174 & 0.4168 \\
    (95\% CI)  & (0.3018, 0.3335) & (0.4042,  0.4292) \\
    \hline
    DBP  & 0.2638 & 0.2726 \\
    (95\% CI)  & (0.2496, 0.2771) & (0.2588, 0.2859) \\
    \hline
    BMI  & 0.0801 & 0.0944  \\
    (95\% CI)  & (0.0676, 0.0926) & (0.0841, 0.1050) \\
    \hline 
    Cholesterol  & -0.0241 & 0.0055 \\
    (95\% CI)  & (-0.0312, -0.0176) & (-0.0024, 0.0134) \\
    \hline 
    HbA1c  & 0.0350 & 0.0834   \\
    (95\% CI)  & (0.0270, 0.0428) & (0.0737, 0.0939) \\
    \hline
    Smoking & 0.6757 & 0.6951 \\
    (95\% CI)  & (0.6585, 0.6933) & (0.6763, 0.7133) \\
    \hline
    \end{tabular}
     \medskip
     \tiny
     \centering
     \medskip
    \\Table 10: Prediction performance for female and male subgroup in test set. \\ For categorical variable prediction AUC shown, for continuous variables $R^{2}$ shown.
\end{table}

Also the prediction performance between the two age groups was evaluated (see table 11). It can be derived that the age level has an influence on prediction performance for the risk factors SBP, DBP, HbA1c, gender and smoking status. Also Gerrits et al. \cite{gerrits2020age} found that age level has an influence on performance for the variables SBP and HbA1c. For these two variables they found a substantial decrease in performance between the age groups 39 to 50 and age above 50. From an $R^{2}$ score of 0.3 to 0.04 for SBP and from 0.24 to 0.06 for HbA1c.

\begin{table}[H]
\centering
 \scriptsize
   \label{tab:table1}
    \begin{tabular}{c|c|c}
     \textbf{Risk Factor} & \textbf{39 < Age <= 50} & \textbf{Age > 50}  \\
    \hline
    SBP  & 0.3660 & 0.3274 \\
    (95\% CI)  & (0.3490, 0.3817) & (0.3135, 0.3417) \\
    \hline
    DBP  & 0.3409 & 0.2611 \\
    (95\% CI)  & (0.3276, 0.3544) & (0.2485, 0.2735) \\
    \hline
    BMI  & 0.0961 & 0.0916  \\
    (95\% CI)  & (0.0836, 0.1070) & (0.0806, 0.1027) \\
    \hline 
    Cholesterol  & 0.0051 & 0.0014 \\
    (95\% CI)  & (-0.0019, 0.0114) & (-0.0040, 0.0067) \\
    \hline 
    HbA1c  & 0.0343 & 0.0161   \\
    (95\% CI)  & (0.0267, 0.0417) & (0.0089, 0.0232) \\
    \hline
    Gender  & 0.9591 & 0.9509 \\
    (95\% CI)  & (0.9559, 0.9621) & (0.9473, 0.9544) \\
    \hline
    Smoking & 0.6495 & 0.6843 \\
    (95\% CI)  & (0.6347, 0.6649) & (0.6651, 0.7034) \\
    \hline
    \end{tabular}
     \medskip
     \tiny
     \centering
     \medskip
    \\Table 11: Prediction performance for age subgroups in test set. \\ For categorical variable prediction AUC shown, for continuous variables $R^{2}$ shown.
\end{table}

Further it can be observed that the age prediction performance of the model decreases at an age of around 65 years (see figure 3), which is in line with the findings of Kim et al. \cite{kim2020effects}, who state that the ageing process observable in the retinal fundus images may saturate at an age of 60 years.

\begin{center}
    \includegraphics[scale=0.2]{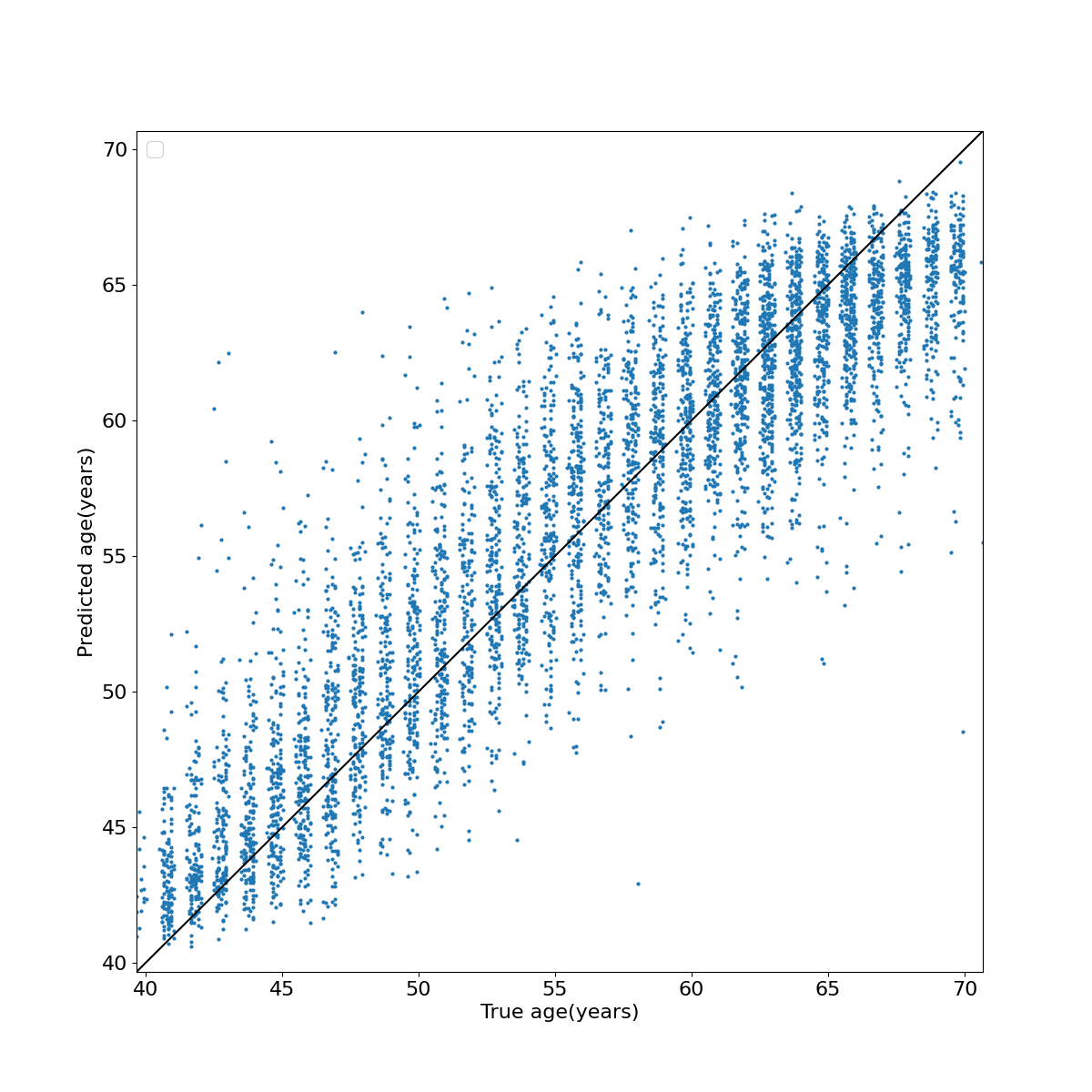}\\
    Figure 3. Actual age (years) vs. predicted age (years)
\end{center}

\section{Discussion}

Concerning the quality assessment model, the results are in line with the findings of Zago et al. \cite{zago2018retinal}, who found that the fine-tuned model with pre-initialized weights outperformed the model where only the last layers' weights were trained. In my analysis the fine-tuned quality assessment model achieved a better performance (AUC of 97.80\%) than the model where only the weights of the last added layers were trained (AUC of 97.07\%). To filter out bad quality images the threshold has been set to 0.5. Future analysis might investigate if by setting a stricter threshold and thereby putting more focus on good quality images leads to an increase in performance. Also, the quality assessment model in my research classified some images that were clearly bad quality (completely dark images) as good quality. Therefore to optimize the model more such dark images need to be labelled and included in the training process of the model. Approximately 60 left retinal images and 60 right retinal images of such clearly bad quality images (circular mask could not be detected) were still included in the CVD risk factor prediction models generation and therefore might have lead to a performance deterioration. Given that there still might be some images of bad quality included in the data set this might be an explanation for the inferior performance compared to the models of Poplin et al. \cite{poplin2018prediction}.

The results of the CVD risk factor prediction models are similar to the findings of Poplin et al. \cite{poplin2018prediction}, but for most CVD risk factors except for age and SBP Poplin et al. \cite{poplin2018prediction} yielded superior prediction results. Even though they used a different training approach by using multi task learning as opposed to single task learning, which I applied in my study, I applied a contrast enhancement technique in the form of gaussian filtering and the combination of left and right retinal image predictions, which are both supposed to increase performance. Poplin et al. \cite{poplin2018prediction} did not preprocess the retinal fundus images in their analysis and derived predictions on image basis rather than on individual basis. For smoking status prediction Vaghefi et al. \cite{vaghefi2019detection} applied contrast enhancement in form of Gaussian filtering, which resulted in an increased prediction performance of an AUC of 0.88 as opposed to 0.71 in the study of Poplin et al. \cite{poplin2018prediction}, which they claimed to be the main reason for the better performance of their model. Similar to the research of Vaghefi et al. \cite{vaghefi2019detection} I could also show that the application of Gaussian filtering on the input images has a positive impact on the prediction performance of the gender prediction model. Since the variables age and SBP showed a superior performance than the model of Poplin et al. \cite{poplin2018prediction} it could be the case that the Gaussian filtering technique had a larger impact on the prediction performance of these two variables. An explanation could be that Gaussian filtering makes blood vessels visible more clearly in the image \cite{lestari2019retinal}, which according to the findings of Poplin et al. \cite{poplin2018prediction} are particularly important for the model to derive predictions for SBP and age. 

An explanation for the performance difference by the smoking status prediction model of Vaghefi et al. \cite{vaghefi2019detection} and my model could be that the gender status in the smoking class (female 47.81\% and male 52.19\%) and non-smoking class (female 56.20\% and male 43.80\%) is fairly balanced in my training set whereas the gender status in the data set used by Vaghefi et al. \cite{vaghefi2019detection} is imbalanced towards the male gender in the smoking class (66\% male gender). Vaghefi et al. \cite{vaghefi2019detection} state that given this imbalance towards the male gender it is possible that the algorithm was sensitive to sex when making predictions about the smoking status of individuals. Since my data set is fairly balanced in respect to gender within the smoking classes and also the prediction performance of smoking status for the male and female class is fairly consistent (see table 10), my model is highly likely not sensitive to gender when deriving predictions about smoking status.

An explanation for the prediction gap for HbA1c between the study of Gerrits et al. \cite{gerrits2020age}, who yielded an $R^{2}$ score of 0.34 when predicting HbA1c as opposed to an $R^{2}$ of 0.0579 in my study could be that the mean age in the UK Biobank data set used in this analysis is relatively high (56.05 years $\pm 8.24$ years) as opposed to 40.6 $\pm 13$ years in the Qatar Biobank data set. Gerrits et al. \cite{gerrits2020age} found that the prediction performance of HbA1c is influenced by age (years <= 30  $R^{2}$ score of -0.14, 30 < years <= 39 $R^{2}$ score of 0.28, 39 < years <=50 years $R^{2}$ score of 0.24, years > 50 $R^{2}$ score of 0.06) with having the highest prediction performance between 30 and 39 and a relatively low prediction performance above 50.

\section{Conclusion}

This paper showed that the prediction performance for all CVD risk factors trained with UK Biobank retinal images was superior to the baseline models. Also for the CVD risk factors total cholesterol and HbA1c the prediction performance was better than the baseline models. Nevertheless, the prediction performance of total cholesterol and HbA1c was still lower compared to the prediction performance of other risk factors with an $R^{2}$ score of 0.0157 and 0.0579 respectively. This indicates that predictions of total cholesterol and HbA1c from retinal fundus images are only of limited value. 

The prediction performance for the models predicting the CVD risk factors age and SBP is superior to the study of Poplin et al. \cite{poplin2018prediction}, whereas the model of Poplin et al. \cite{poplin2018prediction} achieved a better performance for the prediction of the CVD risk factors gender, BMI, DBP, smoking status. Applying Gaussian filtering as preprocessing technique seems to positively impact the model performance, whereas the image quality assessment model needs to be further adjusted to eliminate all bad quality images from the data set. 

Fine-tuning the inception-v3 model to retinal images as opposed to only training the weights of the last added layers has a substantial positive impact on prediction performance and therefore fine-tuning the CNN model proves to be essential when working with retinal images. 

Even though the data set for the analysis is dominated by individuals from the british/irish no overall superior prediction performance for this population group could be observed. To further evaluate this finding and to evaluate the generalization ability of the models, the models need to be tested on external data sets with different ethnicity distributions.   

I could show that the combination of the predictions from left and right retinal images per individual increases the prediction performance for all CVD risk factors analyzed in this study. Further research could be done by evaluating if the combination of predictions of more than one left and right retinal image per individual further increases the prediction performance.  

I could derive that gender has a significant impact on the prediction performance of the CVD risk factors SBP, cholesterol and HbA1c. The age level plays a significant role in the prediction of the CVD risk factors SBP, DBP, HbA1c, gender and smoking status. Given that the relation between variables has, to some extent, an impact on the prediction performance, the co-variables in the underlying data set play an essential role \cite{vaghefi2019detection} and therefore need to be considered to when comparing two models.

%This can be seen as explanation why well functioning algorithms perform poorly when applied on a data set from  a different population with different balancing of these co-variables \cite{vaghefi2019detection}. 

\section{Acknowledgments}
The authors thank the UK Biobank \cite{sudlow2015uk} data access granted through application number 55495.

\newpage
\bibliographystyle{plain}
\bibliography{MyThesis}

\begin{thebibliography}{10}

\bibitem{who}
{Cardiovascular Diseases}.
\newblock
  \url{https://www.who.int/health-topics/cardiovascular-diseases#tab=tab_1}.
\newblock [Online; accessed 06-February-2021].

\bibitem{bock2018improvement}
Sebastian Bock, Josef Goppold, and Martin Wei{\ss}.
\newblock An improvement of the convergence proof of the adam-optimizer.
\newblock {\em arXiv preprint arXiv:1804.10587}, 2018.

\bibitem{cheong2017interventions}
AT~Cheong, SM~Liew, EM~Khoo, NF~Mohd Zaidi, and K~Chinna.
\newblock Are interventions to increase the uptake of screening for
  cardiovascular disease risk factors effective? a systematic review and
  meta-analysis.
\newblock {\em BMC family practice}, 18(1):1--15, 2017.

\bibitem{christopher2018performance}
Mark Christopher, Akram Belghith, Christopher Bowd, James~A Proudfoot,
  Michael~H Goldbaum, Robert~N Weinreb, Christopher~A Girkin, Jeffrey~M
  Liebmann, and Linda~M Zangwill.
\newblock Performance of deep learning architectures and transfer learning for
  detecting glaucomatous optic neuropathy in fundus photographs.
\newblock {\em Scientific reports}, 8(1):1--13, 2018.

\bibitem{dahlof2010cardiovascular}
Bj{\"o}rn Dahl{\"o}f.
\newblock Cardiovascular disease risk factors: epidemiology and risk
  assessment.
\newblock {\em The American journal of cardiology}, 105(1):3A--9A, 2010.

\bibitem{dai2020exploring}
Guangzheng Dai, Wei He, Ling Xu, Eric~E Pazo, Tiezhu Lin, Shasha Liu, and
  Chenguang Zhang.
\newblock Exploring the effect of hypertension on retinal microvasculature
  using deep learning on east asian population.
\newblock {\em PloS one}, 15(3):e0230111, 2020.

\bibitem{ding2014retinal}
Jie Ding, Khin~Lay Wai, Kevin McGeechan, et~al.
\newblock Retinal vascular caliber and the development of hypertension: a
  meta-analysis of individual participant data.
\newblock {\em Journal of hypertension}, 32(2):207, 2014.

\bibitem{esteva2017dermatologist}
Andre Esteva, Brett Kuprel, Roberto~A Novoa, Justin Ko, Susan~M Swetter,
  Helen~M Blau, and Sebastian Thrun.
\newblock Dermatologist-level classification of skin cancer with deep neural
  networks.
\newblock {\em nature}, 542(7639):115--118, 2017.

\bibitem{gerrits2020age}
Nele Gerrits, Bart Elen, Toon Van~Craenendonck, Danai Triantafyllidou,
  Ioannis~N Petropoulos, Rayaz~A Malik, and Patrick De~Boever.
\newblock Age and sex affect deep learning prediction of cardiometabolic risk
  factors from retinal images.
\newblock {\em Scientific reports}, 10(1):1--9, 2020.

\bibitem{gonzalez2019iterative}
Cristina Gonz{\'a}lez-Gonzalo, Bart Liefers, Bram van Ginneken, and Clara~I
  S{\'a}nchez.
\newblock Iterative augmentation of visual evidence for weakly-supervised
  lesion localization in deep interpretability frameworks.
\newblock {\em arXiv preprint arXiv:1910.07373}, 2019.

\bibitem{graham2015kaggle}
Ben Graham.
\newblock Kaggle diabetic retinopathy detection competition report.
\newblock {\em University of Warwick}, 2015.

\bibitem{gulshan2016development}
Varun Gulshan, Lily Peng, Marc Coram, Martin~C Stumpe, Derek Wu, Arunachalam
  Narayanaswamy, Subhashini Venugopalan, Kasumi Widner, Tom Madams, Jorge
  Cuadros, et~al.
\newblock Development and validation of a deep learning algorithm for detection
  of diabetic retinopathy in retinal fundus photographs.
\newblock {\em Jama}, 316(22):2402--2410, 2016.

\bibitem{islam2019identifying}
Kh~Tohidul Islam, Sudanthi Wijewickrema, and Stephen O'Leary.
\newblock Identifying diabetic retinopathy from oct images using deep transfer
  learning with artificial neural networks.
\newblock In {\em 2019 IEEE 32nd International Symposium on Computer-Based
  Medical Systems (CBMS)}, pages 281--286. IEEE, 2019.

\bibitem{kim2020effects}
Yong~Dae Kim, Kyoung~Jin Noh, Seong~Jun Byun, Soochahn Lee, Tackeun Kim,
  Leonard Sunwoo, Kyong~Joon Lee, Si-Hyuck Kang, Kyu~Hyung Park, and Sang~Jun
  Park.
\newblock Effects of hypertension, diabetes, and smoking on age and sex
  prediction from retinal fundus images.
\newblock {\em Scientific reports}, 10(1):1--14, 2020.

\bibitem{kingma2014adam}
Diederik~P Kingma and Jimmy Ba.
\newblock Adam: A method for stochastic optimization.
\newblock {\em arXiv preprint arXiv:1412.6980}, 2014.

\bibitem{lecun2015deep}
Yann LeCun, Yoshua Bengio, and Geoffrey Hinton.
\newblock Deep learning.
\newblock {\em nature}, 521(7553):436--444, 2015.

\bibitem{lestari2019retinal}
T~Lestari, A~Luthfi, et~al.
\newblock Retinal blood vessel segmentation using gaussian filter.
\newblock In {\em Journal of Physics: Conference Series}, volume 1376, page
  012023. IOP Publishing, 2019.

\bibitem{li2018automated}
Zhixi Li, Stuart Keel, Chi Liu, Yifan He, Wei Meng, Jane Scheetz, Pei~Ying Lee,
  Jonathan Shaw, Daniel Ting, Tien~Yin Wong, et~al.
\newblock An automated grading system for detection of vision-threatening
  referable diabetic retinopathy on the basis of color fundus photographs.
\newblock {\em Diabetes care}, 41(12):2509--2516, 2018.

\bibitem{lopez2013insight}
Victoria L{\'o}pez, Alberto Fern{\'a}ndez, Salvador Garc{\'\i}a, Vasile Palade,
  and Francisco Herrera.
\newblock An insight into classification with imbalanced data: Empirical
  results and current trends on using data intrinsic characteristics.
\newblock {\em Information sciences}, 250:113--141, 2013.

\bibitem{macgillivray2015suitability}
Thomas~J MacGillivray, James~R Cameron, Qiuli Zhang, Ahmed El-Medany, Carl
  Mulholland, Ziyan Sheng, Bal Dhillon, Fergus~N Doubal, Paul~J Foster,
  Emmanuel Trucco, et~al.
\newblock Suitability of uk biobank retinal images for automatic analysis of
  morphometric properties of the vasculature.
\newblock {\em PLoS One}, 10(5):e0127914, 2015.

\bibitem{mcgeechan2009meta}
Kevin McGeechan, Gerald Liew, Petra Macaskill, Les Irwig, Ronald Klein,
  Barbara~EK Klein, Jie~Jin Wang, Paul Mitchell, Johannes~R Vingerling,
  Paulus~TVM DeJong, et~al.
\newblock Meta-analysis: retinal vessel caliber and risk for coronary heart
  disease.
\newblock {\em Annals of internal medicine}, 151(6):404--413, 2009.

\bibitem{mookiah2013computer}
Muthu Rama~Krishnan Mookiah, U~Rajendra Acharya, Chua~Kuang Chua, Choo~Min Lim,
  EYK Ng, and Augustinus Laude.
\newblock Computer-aided diagnosis of diabetic retinopathy: A review.
\newblock {\em Computers in biology and medicine}, 43(12):2136--2155, 2013.

\bibitem{nixon2010non}
Richard~M Nixon, David Wonderling, and Richard~D Grieve.
\newblock Non-parametric methods for cost-effectiveness analysis: the central
  limit theorem and the bootstrap compared.
\newblock {\em Health economics}, 19(3):316--333, 2010.

\bibitem{perez2017effectiveness}
Luis Perez and Jason Wang.
\newblock The effectiveness of data augmentation in image classification using
  deep learning.
\newblock {\em arXiv preprint arXiv:1712.04621}, 2017.

\bibitem{poplin2018prediction}
Ryan Poplin, Avinash~V Varadarajan, Katy Blumer, Yun Liu, Michael~V McConnell,
  Greg~S Corrado, Lily Peng, and Dale~R Webster.
\newblock Prediction of cardiovascular risk factors from retinal fundus
  photographs via deep learning.
\newblock {\em Nature Biomedical Engineering}, 2(3):158--164, 2018.

\bibitem{rim2020prediction}
Tyler~Hyungtaek Rim, Geunyoung Lee, Youngnam Kim, Yih-Chung Tham, Chan~Joo Lee,
  Su~Jung Baik, Young~Ah Kim, Marco Yu, Mihir Deshmukh, Byoung~Kwon Lee, et~al.
\newblock Prediction of systemic biomarkers from retinal photographs:
  development and validation of deep-learning algorithms.
\newblock {\em The Lancet Digital Health}, 2(10):e526--e536, 2020.

\bibitem{ruder2016overview}
Sebastian Ruder.
\newblock An overview of gradient descent optimization algorithms.
\newblock {\em arXiv preprint arXiv:1609.04747}, 2016.

\bibitem{russakovsky2015imagenet}
Olga Russakovsky, Jia Deng, Hao Su, Jonathan Krause, Sanjeev Satheesh, Sean Ma,
  Zhiheng Huang, Andrej Karpathy, Aditya Khosla, Michael Bernstein, et~al.
\newblock Imagenet large scale visual recognition challenge.
\newblock {\em International journal of computer vision}, 115(3):211--252,
  2015.

\bibitem{sam2019offline}
Suriani~Mohd Sam, Kamilia Kamardin, Nilam Nur~Amir Sjarif, Norliza Mohamed,
  et~al.
\newblock Offline signature verification using deep learning convolutional
  neural network (cnn) architectures googlenet inception-v1 and inception-v3.
\newblock {\em Procedia Computer Science}, 161:475--483, 2019.

\bibitem{shin2016deep}
Hoo-Chang Shin, Holger~R Roth, Mingchen Gao, Le~Lu, Ziyue Xu, Isabella Nogues,
  Jianhua Yao, Daniel Mollura, and Ronald~M Summers.
\newblock Deep convolutional neural networks for computer-aided detection: Cnn
  architectures, dataset characteristics and transfer learning.
\newblock {\em IEEE transactions on medical imaging}, 35(5):1285--1298, 2016.

\bibitem{srivastava2014dropout}
Nitish Srivastava, Geoffrey Hinton, Alex Krizhevsky, Ilya Sutskever, and Ruslan
  Salakhutdinov.
\newblock Dropout: a simple way to prevent neural networks from overfitting.
\newblock {\em The journal of machine learning research}, 15(1):1929--1958,
  2014.

\bibitem{sudlow2015uk}
Cathie Sudlow, John Gallacher, Naomi Allen, Valerie Beral, Paul Burton, John
  Danesh, Paul Downey, Paul Elliott, Jane Green, Martin Landray, et~al.
\newblock Uk biobank: an open access resource for identifying the causes of a
  wide range of complex diseases of middle and old age.
\newblock {\em PLoS medicine}, 12(3):e1001779, 2015.

\bibitem{szegedy2016rethinking}
Christian Szegedy, Vincent Vanhoucke, Sergey Ioffe, Jon Shlens, and Zbigniew
  Wojna.
\newblock Rethinking the inception architecture for computer vision.
\newblock In {\em Proceedings of the IEEE conference on computer vision and
  pattern recognition}, pages 2818--2826, 2016.

\bibitem{tajbakhsh2016convolutional}
Nima Tajbakhsh, Jae~Y Shin, Suryakanth~R Gurudu, R~Todd Hurst, Christopher~B
  Kendall, Michael~B Gotway, and Jianming Liang.
\newblock Convolutional neural networks for medical image analysis: Full
  training or fine tuning?
\newblock {\em IEEE transactions on medical imaging}, 35(5):1299--1312, 2016.

\bibitem{ting2018eyeing}
Daniel Shu~Wei Ting and Tien~Yin Wong.
\newblock Eyeing cardiovascular risk factors.
\newblock {\em Nature biomedical engineering}, 2(3):140--141, 2018.

\bibitem{vaghefi2019detection}
Ehsan Vaghefi, Song Yang, Sophie Hill, Gayl Humphrey, Natalie Walker, and David
  Squirrell.
\newblock Detection of smoking status from retinal images; a convolutional
  neural network study.
\newblock {\em Scientific reports}, 9(1):1--9, 2019.

\bibitem{welikala2016automated}
RA~Welikala, MM~Fraz, PJ~Foster, PH~Whincup, Alicja~R Rudnicka, Christopher~G
  Owen, DP~Strachan, Sarah~A Barman, et~al.
\newblock Automated retinal image quality assessment on the uk biobank dataset
  for epidemiological studies.
\newblock {\em Computers in biology and medicine}, 71:67--76, 2016.

\bibitem{whitehill2016exploiting}
Jacob Whitehill.
\newblock Exploiting an oracle that reports auc scores in machine learning
  contests.
\newblock In {\em Proceedings of the AAAI Conference on Artificial
  Intelligence}, volume~30, 2016.

\bibitem{yu2017deep}
Yuhai Yu, Hongfei Lin, Jiana Meng, Xiaocong Wei, Hai Guo, and Zhehuan Zhao.
\newblock Deep transfer learning for modality classification of medical images.
\newblock {\em Information}, 8(3):91, 2017.

\bibitem{zago2018retinal}
Gabriel~Tozatto Zago, Rodrigo~Varej{\~a}o Andre{\~a}o, Bernadette Dorizzi, and
  Evandro Ottoni~Teatini Salles.
\newblock Retinal image quality assessment using deep learning.
\newblock {\em Computers in biology and medicine}, 103:64--70, 2018.

\bibitem{zhang2020prediction}
Li~Zhang, Mengya Yuan, Zhen An, Xiangmei Zhao, Hui Wu, Haibin Li, Ya~Wang,
  Beibei Sun, Huijun Li, Shibin Ding, et~al.
\newblock Prediction of hypertension, hyperglycemia and dyslipidemia from
  retinal fundus photographs via deep learning: A cross-sectional study of
  chronic diseases in central china.
\newblock {\em PloS one}, 15(5):e0233166, 2020.

\bibitem{zhou2019predictive}
Yiwang Zhou, Lu~Zhao, Nina Zhou, Yi~Zhao, Simeone Marino, Tuo Wang, Hanbo Sun,
  Arthur~W Toga, and Ivo~D Dinov.
\newblock Predictive big data analytics using the uk biobank data.
\newblock {\em Scientific reports}, 9(1):1--10, 2019.

\end{thebibliography}

\newpage
\newpage
\appendix

\section{CVD Risk Factor Prediction Models - Training of last layers}

\begin{center}

\begin{table}[H]
\centering
 \scriptsize
   \label{tab:table1}
    \begin{tabular}{c|c|c|c}
     \textbf{Risk Factor} & \textbf{Prediction Basis} & \textbf{AUC} &  \textbf{Baseline}\\
     \hline
    Gender   & L+R Images & 0.6639   & 0.5 \\
    (95\% CI) & & (0.66, 0.67) & \\
    \hline
    Gender & L Images & 0.6690 &  0.5  \\
    
    (95\% CI) & & (0.66, 0.68) &  \\
    
    \hline
    Gender & R Images & 0.6589   & 0.5 \\
    
    (95\% CI) & & (0.65, 0.67) & \\
    
    \hline
    Gender & Average L+R  &  0.6701 & 0.5 \\
    (95\% CI) & Image &  (0.66, 0.68) &    \\
    \hline
    
    \end{tabular}
     \medskip
     \tiny
     \centering
     \medskip
    \\Table 12: Gender prediction with inception-v3 model with training of weights of last layers

\end{table}

\vspace{-2cm}

\begin{table}[H]
\centering
 \scriptsize
   \label{tab:table1}
    \begin{tabular}{c|c|c|c|c|c}
     \textbf{Risk Factor} & \textbf{Prediction Basis} & \textbf{MAE} & \textbf{R2 Score} & \textbf{Baseline MAE} & \textbf{Baseline R2}\\
     \hline
    Age   & L+R Images & 4.8692   & 0.4575 & 7.1681 & 0\\
    (95\% CI) & & (4.80, 4.93) & (0.44, 0.47) & & \\
    \hline
    Age & L Images & 4.8757 & 0.4543 & 7.1681 & 0 \\
    
    (95\% CI) & & (4.80, 4.93) & (0.44, 0.47) & & \\
\hline
    
    Age & R Images & 4.8626 & 0.4607 & 7.1681 & 0 \\
    
    (95\% CI) & & (4.80, 4.92) & (0.45, 0.47) & & \\
    
  \hline  
    Age & Average L+R  & 4.6029 & 0.5220 & 7.1681 & 0  \\
    (95\% CI) & Image &  (4.54, 4.66) & (0.51, 0.53) & &\\
    \hline
    
    \end{tabular}
     \medskip
     \tiny
     \centering
     \medskip
    \\Table 13: Age prediction with inception-v3 model with training of weights of last layers
    
\end{table}

\end{center}

\section{CVD Risk Factor Prediction Models - Training of all layers}

\begin{table}[H]
\centering
 \scriptsize
   \label{tab:table1}
    \begin{tabular}{c|c|c|c|c}
     \textbf{Risk Factor} & \textbf{Prediction Basis} & \textbf{AUC} & \textbf{AUC-PR} & \textbf{Baseline}\\
     \hline
    Gender   & L+R Images & 0.9291 & 0.9170 & 0.5 \\
    (95\% CI) & & (0.92, 0.93) & (0.91, 0.92)&  \\
    \hline
    Gender & L Images & 0.9307   & 0.9185 & 0.5  \\
    (95\% CI) & & (0.93, 0.94) & (0.91, 0.92)& \\
    \hline
    Gender & R Images & 0.9286  & 0.9167 & 0.5 \\
    (95\% CI) & & (0.92, 0.93)& (0.91, 0.92) &\\
    \hline
    Gender & Average L+R  & 0.9533 & 0.9458 & 0.5  \\
    (95\% CI) & Image &  (0.95, 0.96)  & (0.94, 0.95) & \\
    \hline
    
    Smoking Undersamp. & L+R Images & 0.5600   & 0.1100 & 0.5 \\
    (95\% CI) & & (0.54, 0.58) & (0.10, 0.12) & \\
    \hline
    Smoking Undersamp. & L Images & 0.5542   & 0.1085 & 0.5 \\
    (95\% CI) & & (0.54, 0.57) & (0.10, 0.12) & \\
    \hline
    Smoking Undersamp. & R Images & 0.5667  & 0.1136 & 0.5 \\
    (95\% CI) & & (0.55, 0.58) &(0.10, 0.12) & \\
    \hline
    Smoking Undersamp. & Average L+R  & 0.5666   & 0.1156 & 0.5  \\
    (95\% CI) & Image &  (0.55, 0.58) & (0.11, 0.13) & \\
    \hline
    
    Smoking Weighted Loss & L+R Images & 0.6716  &0.2050 & 0.5   \\
    (95\% CI) & & (0.65, 0.69) & (0.18, 0.23) & \\
    \hline 
    Smoking Weighted Loss & L Images & 0.6680   & 0.1968 & 0.5 \\
    (95\% CI) & & (0.65, 0.69) & (0.18, 0.22) & \\
    \hline
    Smoking Weighted Loss & R Images & 0.6754  & 0.212 & 0.5 \\
    (95\% CI) & & (0.66, 0.69)& (0.19, 0.24) & \\
    \hline
    Smoking Weighted Loss & Average L+R  & 0.6856  & 0.2298 & 0.5  \\
    (95\% CI) & Image &  (0.67, 0.70) & (0.21, 0.26) & \\
    \hline
    
    \end{tabular}
     \medskip
     \tiny
     \centering
     \medskip
    \\Table 14: Categorical risk factor prediction with inception-v3 model with fine-tuning of all layers
\end{table}

\begin{table}[H]
\centering
 \scriptsize
   \label{tab:table1}
    \begin{tabular}{c|c|c|c|c|c}
    \textbf{Risk Factor} & \textbf{Prediction Basis} & \textbf{MAE} & \textbf{R2 Score} & \textbf{Baseline} & \textbf{Baseline}\\
    \textbf{} & \textbf{} & \textbf{} & \textbf{} & \textbf{MAE} & \textbf{R2}\\
    \hline
    Age  & L+R Images & 2.9996   & 0.7714 & 7.1681 & 0\\
    (95\% CI) & & (2.96, 3.04) & (0.76, 0.78) & & \\
    \hline
    Age & L Images & 3.0233 & 0.7667 & 7.1681 & 0 \\
    (95\% CI) & & (2.98, 3.07) & (0.76, 0.78) & & \\

    \hline
    Age & R Images & 2.9759 & 0.7761 & 7.1681 & 0 \\
    (95\% CI) & & (2.93, 3.02) & (0.77, 0.78) & & \\
    
    \hline
    Age & Average L+R  & 2.7711 & 0.8070 & 7.1681 & 0  \\
    (95\% CI) & Image &  (2.73, 2.81) & (0.80, 0.81) & &\\
    \hline
    SBP   & L+R Images & 11.5811   & 0.3303 & 14.3644 & 0\\
    (95\% CI) & & (11.42, 11.75) & (0.31, 0.34) & & \\
    \hline
    SBP & L Images & 11.7272 & 0.3203 & 14.3644  & 0 \\
    (95\% CI) & & (11.56, 11.89) & (0.31, 0.34) & & \\

    \hline
    SBP & R Images & 11.4349 & 0.3402 & 14.3644  & 0 \\
    (95\% CI) & & (11.26, 11.60) & (0.32, 0.36) & & \\
    
    \hline
    SBP & Average L+R  & 11.0036 & 0.3927 & 14.3644  & 0  \\
    (95\% CI) & Image &  (10.85, 11.16) & (0.38, 0.41) & &\\
    \hline
    DBP   & L+R Images & 6.8615   & 0.2499 & 7.9256 & 0\\
    (95\% CI) & & (6.77, 6.96) & (0.24, 0.26) & & \\
    \hline
    DBP & L Images & 6.8745 & 0.2441 & 7.9256 & 0 \\
    (95\% CI) & & (6.78, 6.97) & (0.23, 0.26) & & \\
    \hline
    DBP & R Images & 6.8484 & 0.2557 & 7.9256 & 0 \\
    (95\% CI) & & (6.76, 6.94) & (0.24, 0.27) & & \\
    \hline
    DBP & Average L+R  & 6.6689 & 0.2910 & 7.9256 & 0  \\
    (95\% CI) & Image &  (6.58, 6.76) & (0.28, 0.30) & &\\
    \hline
    
    BMI   & L+R Images & 3.2795   & 0.0675 & 3.4936 & 0\\
    (95\% CI) & & (3.23, 3.33) & (0.05, 0.08) & & \\
    \hline
    BMI & L Images & 3.2837 & 0.0775 & 3.4936 & 0 \\
    (95\% CI) & & (3.23, 3.34) & (0.06, 0.09) & & \\
    \hline
    BMI & R Images & 3.2753 & 0.0576 & 3.4936 & 0 \\
    (95\% CI) & & (3.22, 3.33) & (0.04, 0.07) & & \\
    \hline
    BMI & Average L+R  & 3.2119 & 0.0960 & 3.4936 & 0  \\
    (95\% CI) & Image &  (3.16, 3.27) & (0.08, 0.11) & &\\
    \hline
    Cholesterol   & L+R Images & 0.8893   & 0.0081 & 0.8942 & 0\\
    (95\% CI) & & (0.88, 0.90) & (0.002, 0.014) & & \\
    \hline
    Cholesterol & L Images & 0.8894 & 0.0069 & 0.8942 & 0 \\
    (95\% CI) & & (0.88, 0.90) & (0.0003, 0.013) & & \\
    \hline
    Cholesterol & R Images & 0.8892 & 0.0094 & 0.8942 & 0 \\
    (95\% CI) & & (0.88, 0.90) & (0.003, 0.015) & & \\
    
    \hline
    Cholesterol & Average L+R  & 0.8858 & 0.0157 & 0.8942 & 0  \\
    (95\% CI) & Image &  (0.87, 0.90) & (0.010, 0.021) & &\\
    \hline
    HbA1c   & L+R Images & 3.4093   & 0.0554 & 3.4936 & 0\\
    (95\% CI) & & (3.32, 3.50) & (0.05, 0.06) & & \\
    \hline
    
    HbA1c & L Images & 3.4115 & 0.0539 & 3.6141 & 0 \\
    (95\% CI) & & (3.33, 3.50) & (0.05, 0.06) & & \\

    \hline
    HbA1c & R Images & 3.4072 & 0.0569 & 3.6141 & 0 \\
    (95\% CI) & & (3.32, 3.49) & (0.05, 0.07) & & \\
   
    \hline
    HbA1c & Average L+R  & 3.3937 & 0.0579 & 3.6141 & 0  \\
    (95\% CI) & Image &  (3.31, 3.48) & (0.05, 0.07) & & \\
    
    \hline
    
    \end{tabular}
     \medskip
     \tiny
     \centering
     \medskip
    \\Table 15: Continuous risk factor prediction with inception-v3 model with fine tuning of all layers
\end{table}

\newpage

\section{Predicted CVD risk factors grouped by gender}

\begin{center}
    \includegraphics[scale=0.3]{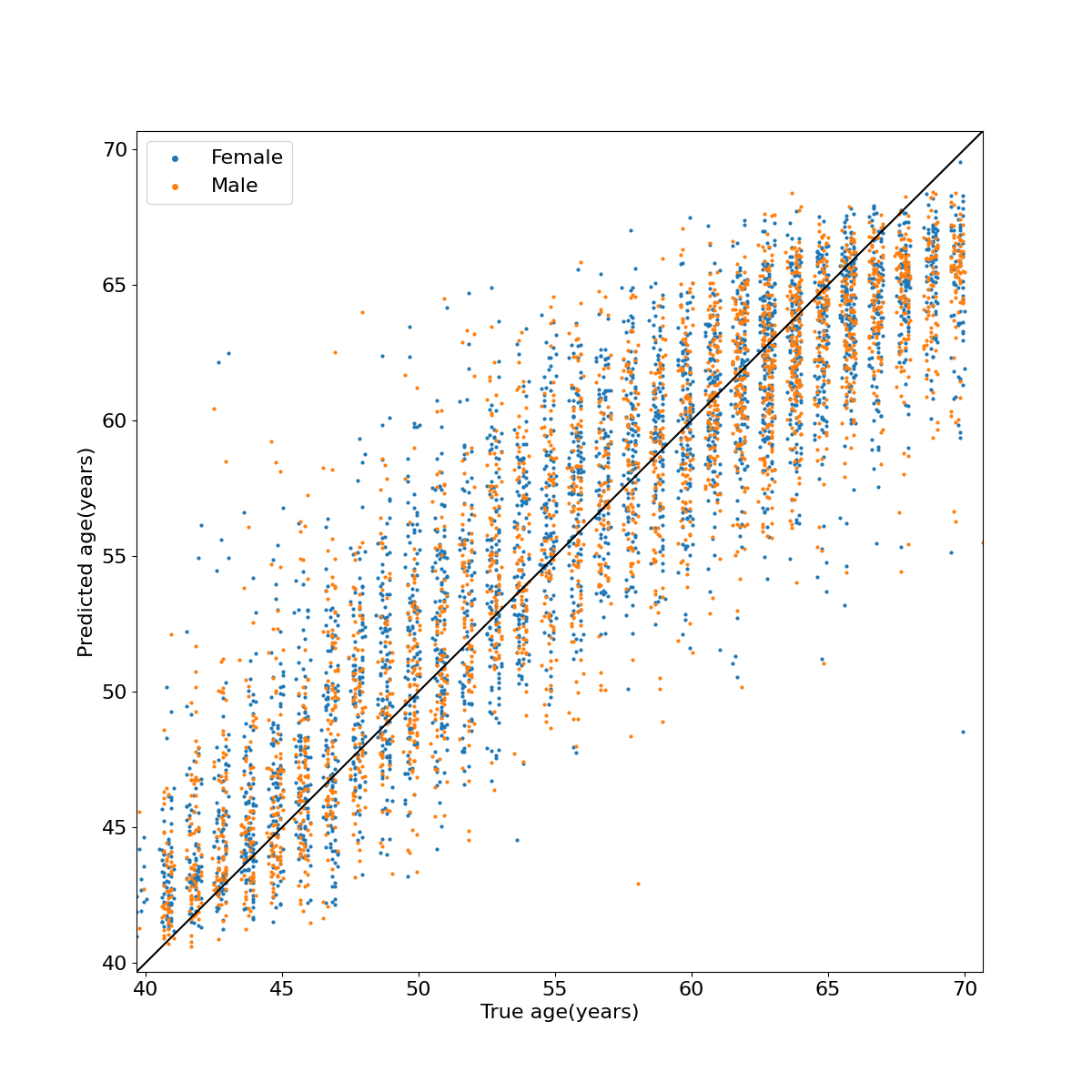}

    \includegraphics[scale=0.3]{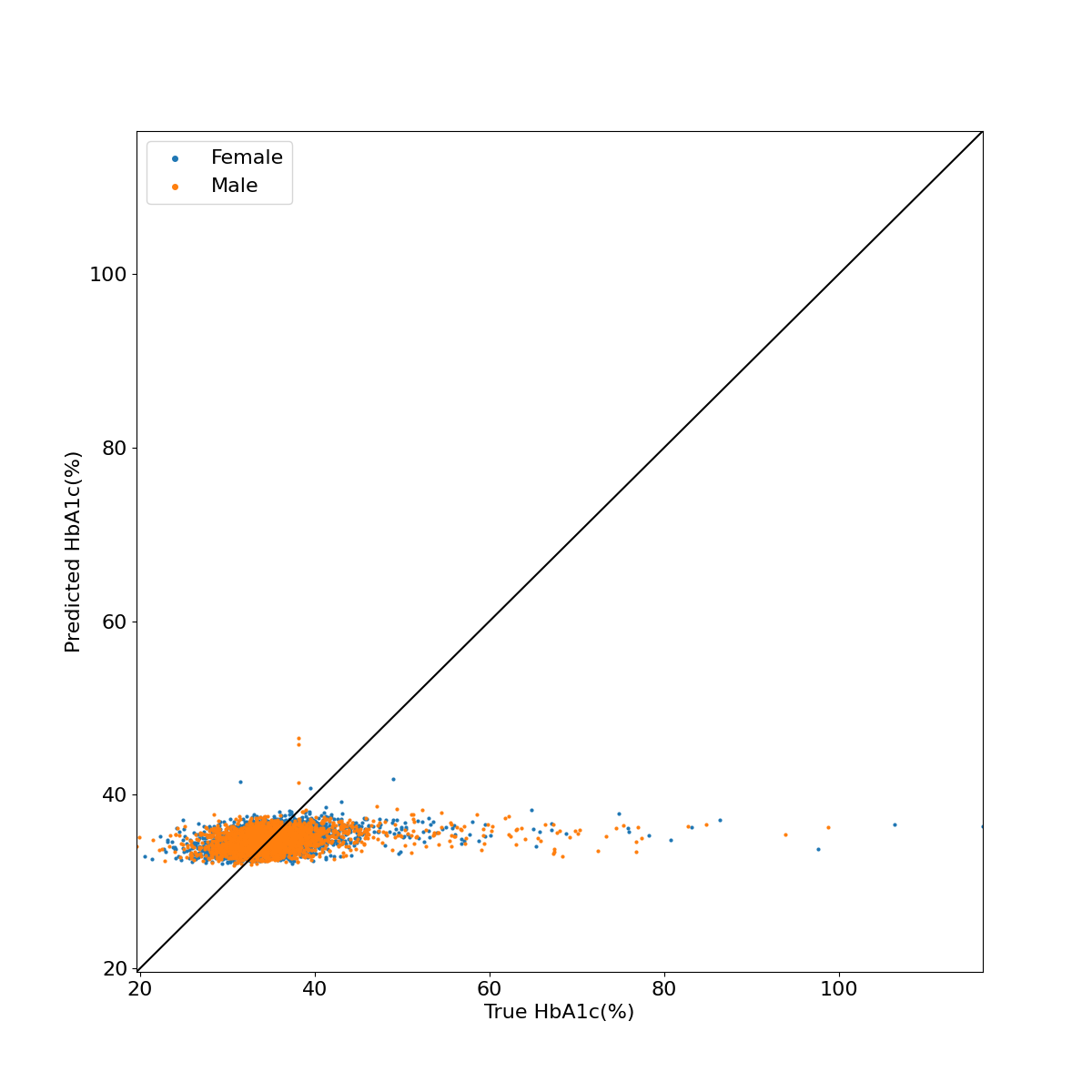}
    
    \includegraphics[scale=0.3]{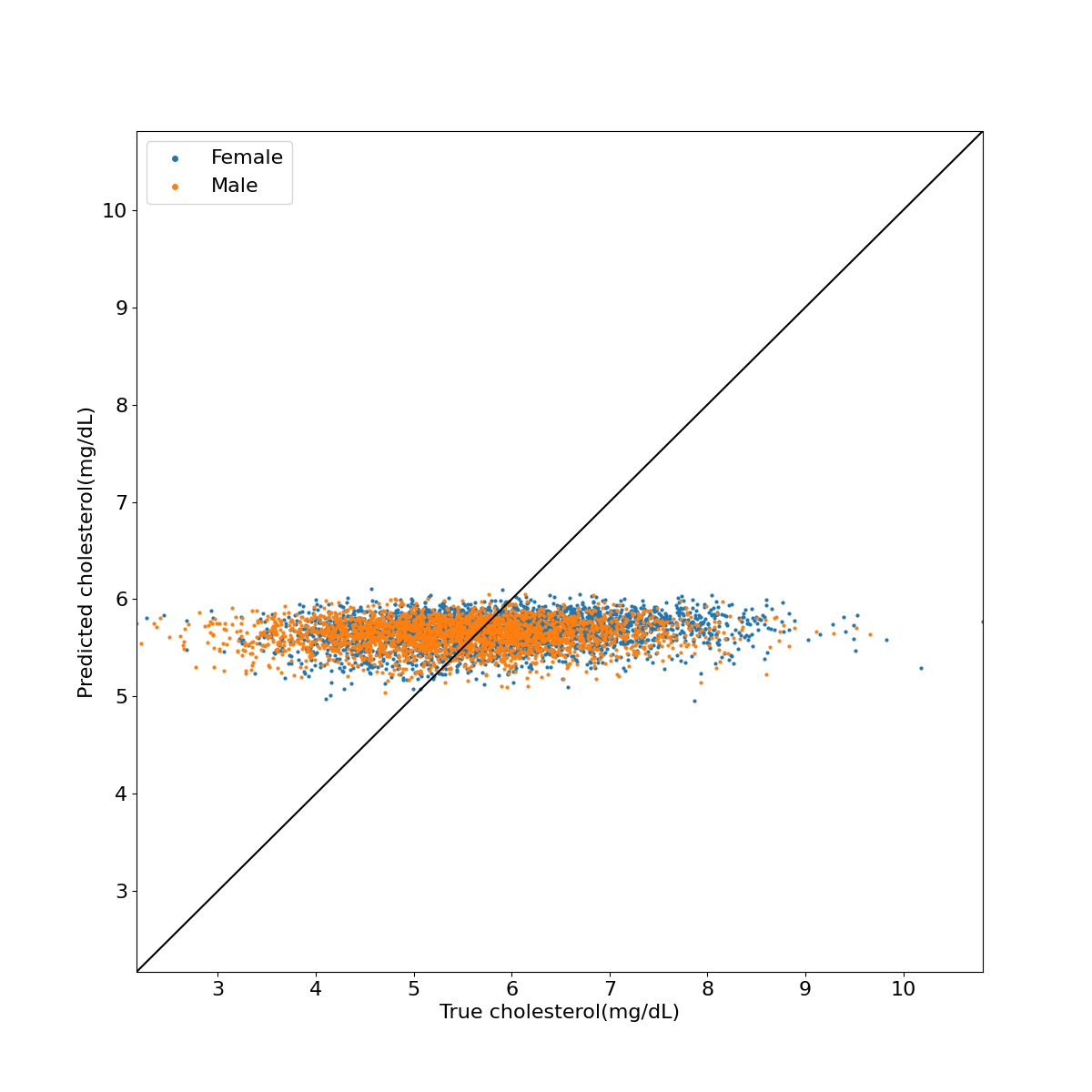}
    
    \includegraphics[scale=0.3]{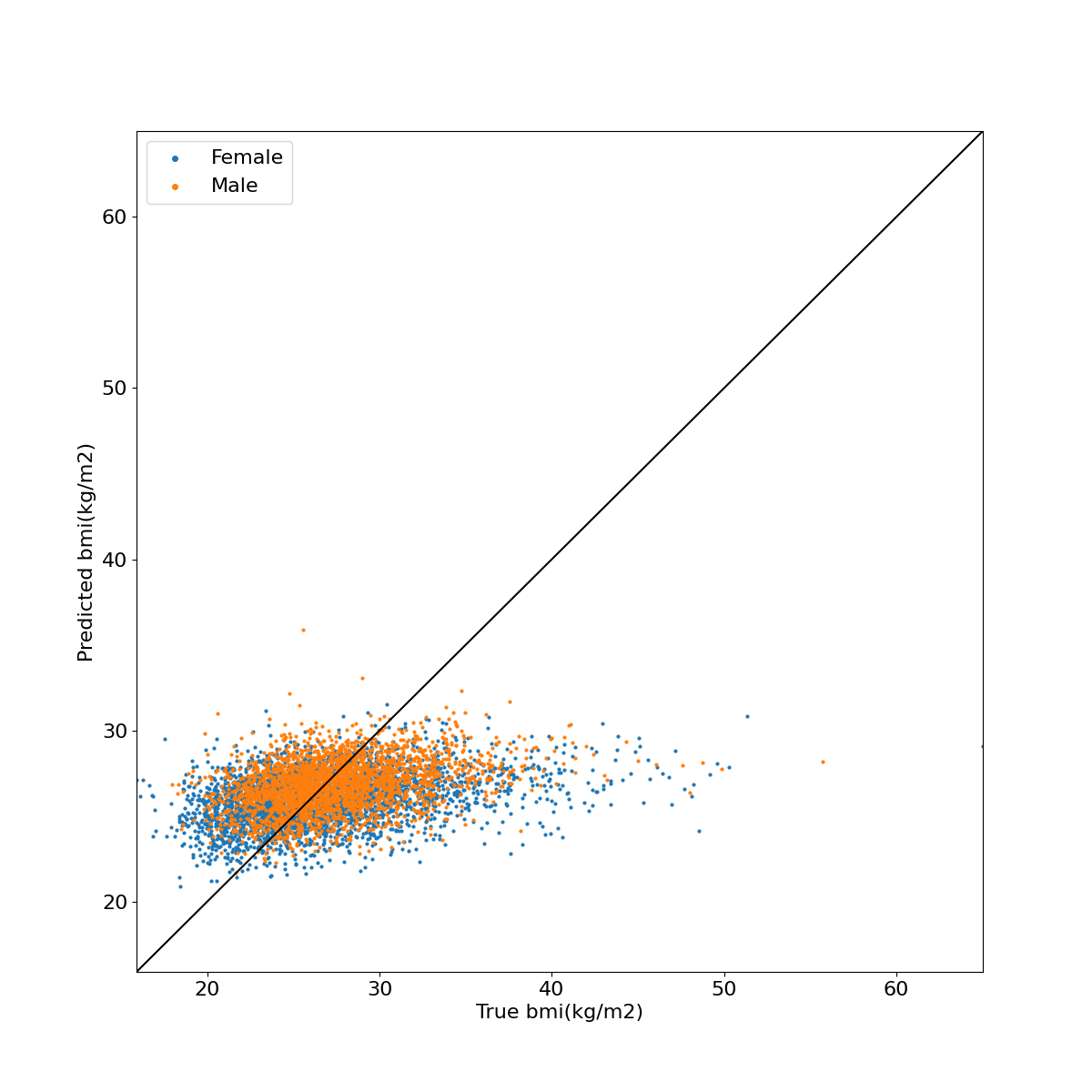}
    
    \includegraphics[scale=0.3]{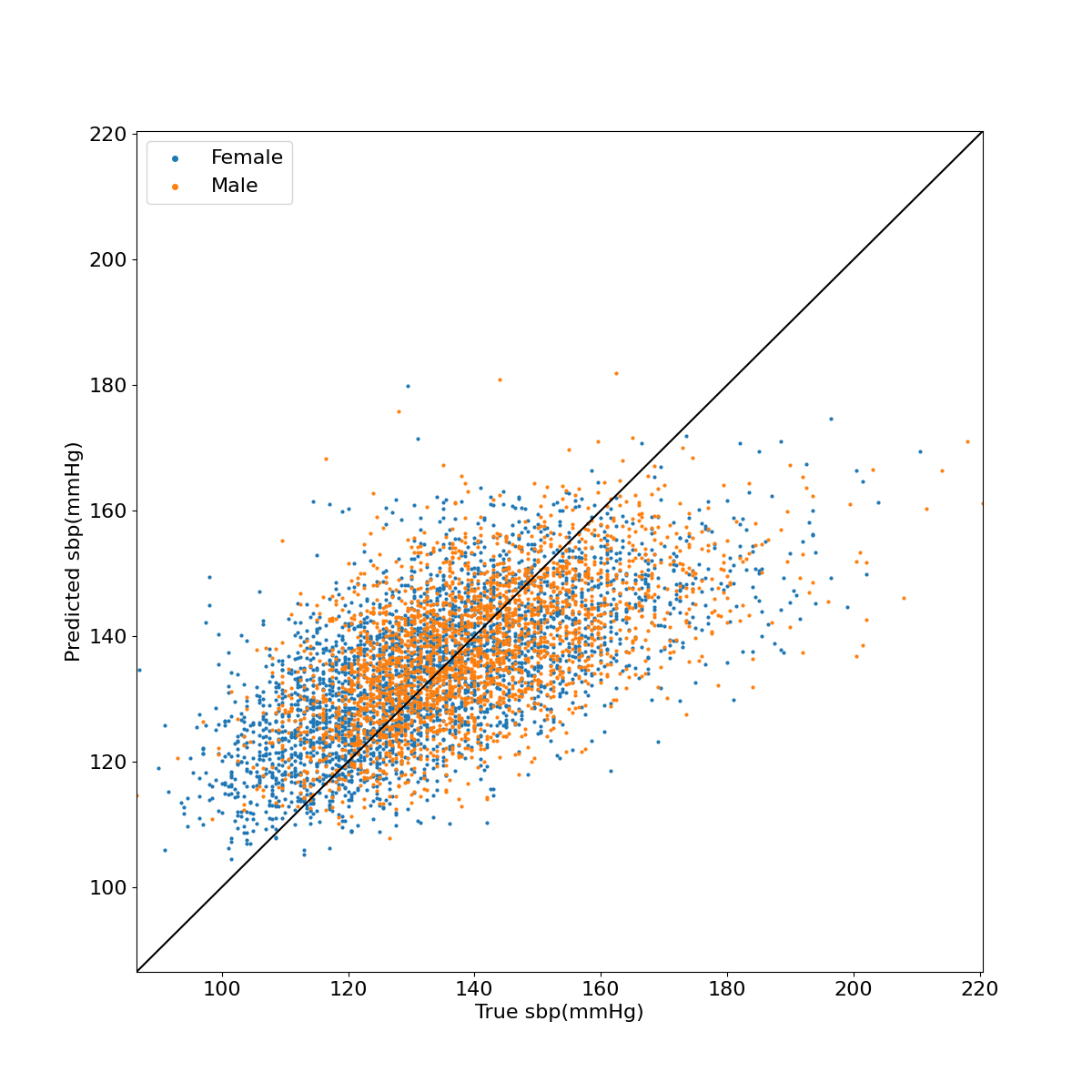}
    
    \includegraphics[scale=0.3]{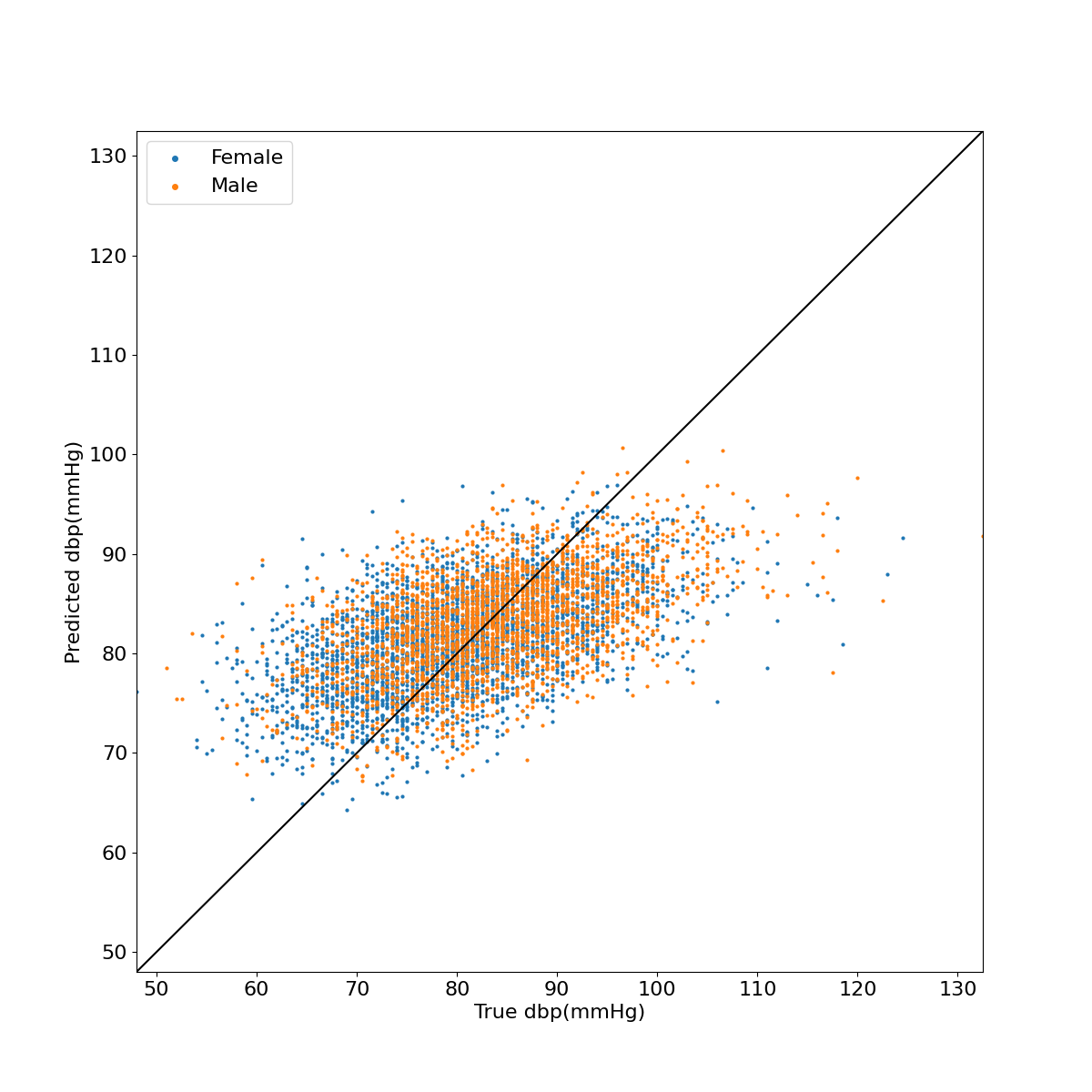}

\end{center}

\end{document}